\documentclass[12pt,amsmath,amssymb,nofootinbib]{revtex4-1}
\usepackage{amsmath} %
\usepackage{pstricks}
\usepackage{color} 
\usepackage{amssymb}
\usepackage{slashed}
\usepackage{graphicx,amsfonts,layout,appendix,subfigure}
\input{epsf} 
\def\beq{\begin{equation}} \def\eeq{\end{equation}}
\def\beqn{\begin{eqnarray}} \def\eeqn{\end{eqnarray}}
\def\bom#1{{\mbox{\boldmath $#1$}}} \def\to{\rightarrow}
\def\nn{\nonumber}

\def\li#1{\mathrm{Li_2}\left(#1\right)}

\def\ln#1{\mathrm{log}\left(#1\right)}
\def\Eq#1{Eq.~(\ref{#1})}

\newcommand\KerLOA{\la {\cal{P}}^{(0)}_{q_1 \bar{q}_2 \gamma_3} \left.\right|_{\epsilon^0} \ra}

\newcommand\IDC{\textbf{Id}}
\newcommand\SUNT{\textit{\textbf{T}}}

\newcommand\DST{D_{\rm ST}}

\newcommand\IC{\textbf{\textit{I}}}

\newcommand\Spamplitude{{\rm Sp}}

\newcommand\CG{c_{\Gamma}}

\newcommand\as{\alpha_{\mathrm{S}}}
\newcommand\g{g_{\mathrm{S}}}

\def\ep{\epsilon}

\def\wp{\widetilde P}

\def\beq{\begin{equation}}
\def\eeq{\end{equation}}
\def\beeq{\begin{eqnarray}}
\def\eeeq{\end{eqnarray}}

\def\bom#1{{\mbox{\boldmath $#1$}}}
\def\to{\rightarrow}

\newcommand{\la}{\langle}
\newcommand{\ra}{\rangle}

\def\nn{\nonumber}

\def\ID{1 \kern -.45 em 1}
\def\RS{{\scriptscriptstyle\rm R\!.S\!.}}
\def\cdr{{\scriptscriptstyle\rm C\!.D\!.R\!.}}
\def\dimr{{\scriptscriptstyle\rm D\!.R\!.}}

\def\sp{{\bom {Sp}}}
\def\ket#1{|{#1}\ra}
\def\bra#1{\la{#1}|}

\def\cbet0{b_0}

\begin{document}

\begin{flushright}
LPN14-100\\ IFIC/14-55
\end{flushright}

\title{Polarized triple-collinear splitting functions\\[1ex] 
at NLO for processes with photons}

\author{Germ\'an F. R. Sborlini~$^{(a, b)}$} \email{gfsborlini@df.uba.ar}
\author{Daniel de Florian~$^{(a)}$} \email{deflo@df.uba.ar}  
\author{Germ\'an Rodrigo~$^{(b)}$}  \email{german.rodrigo@csic.es}

\affiliation{
${}^{(a)}$Departamento de F\'\i sica and IFIBA, FCEyN, Universidad de Buenos Aires, 
(1428) Pabell\'on 1 Ciudad Universitaria, Capital Federal, Argentina \\
${}^{(b)}$Instituto de F\'{\i}sica Corpuscular, 
Universitat de Val\`encia - Consejo Superior de Investigaciones Cient\'{\i}ficas,
Parc Cient\'{\i}fic, E-46980 Paterna, Valencia, Spain}

\date{\today}

\begin{abstract}
We compute the polarized splitting functions in the triple collinear limit at next-to-leading order accuracy (NLO) in the strong coupling $\as$, for the splitting processes $\gamma \to q \bar{q} \gamma$, $\gamma \to q \bar{q} g$ and $g \to q \bar{q} \gamma$. The divergent structure of each splitting function was compared to the predicted behaviour according to Catani's formula. The results obtained in this paper are compatible with the unpolarized splitting functions computed in a previous article. Explicit results for NLO corrections are presented in the context of conventional dimensional regularization (CDR).
\end{abstract}

\maketitle

%***********************************************************************************************************************************************%

%***********************************************************************************************************************************************%
\section{Introduction}
%***********************************************************************************************************************************************%
The multiple-collinear limit of scattering amplitudes in gauge theories is relevant for many reasons. From a phenomenological point of view, higher-order splitting functions are an essential ingredient of subtraction-like algorithms for computing physical cross sections~\cite{Catani:1996vz}. In particular, multiple collinear splitting functions at loop level are required to achieve next-to-next-to-leading order (NNLO) or even higher perturbative orders. Besides that, parton shower (PS) generators make an extensive use of the collinear behaviour of matrix elements. In order to have a complete description of the collinear splitting, it is important to keep spin correlations from the parent parton. This is the main motivation for computing polarized splitting functions at higher-orders, both increasing the number of collinear particles and loops.

Collinear factorization properties~\cite{Catani:2011st} establish that the divergent behaviour of scattering amplitudes is isolated into universal factors called splitting amplitudes~\cite{Berends:1987me, Mangano:1990by}.\footnote{See also Ref. \cite{Collins:1989gx} and references therein.} Besides these well-known properties, strict collinear factorization could be broken in certain kinematic configurations \cite{Catani:2011st,Forshaw:2012bi}. These effects are originated by non-vanishing color correlations among collinear and non-collinear partons, and they could become manifest in the multiple collinear limit at loop-level. So, this constitutes another motivation for exploring higher-order corrections to polarized splitting functions with more than two collinear partons.

For the double-collinear limit at the level of squared matrix-elements, splitting functions are usually called Altarelli-Parisi (AP) kernels~\cite{Altarelli:1977zs}. They have been computed at 
one-loop~\cite{Bern:1998sc, Bern:1999ry, Bern:1993qk, Bern:1994zx,Bern:1995ix,Kosower:1999rx,Sborlini:2013jba} 
and two-loop level~\cite{Bern:2004cz, Badger:2004uk, Vogt:2005dw, Vogt:2004mw, Moch:2004pa}, both for amplitudes and squared matrix-elements. For the multiple collinear limit, tree-level splitting functions were computed by several authors~\cite{Campbell:1997hg, Catani:1998nv, Catani:1999ss, DelDuca:1999ha, Birthwright:2005ak, Birthwright:2005vi}. Although a full one-loop description is still missing, there are some specific results for the triple collinear limit of one-loop amplitudes 
for the antisymmetric part of $q \to q Q \bar{Q}$~\cite{Catani:2003vu} and for processes involving at least one photon \cite{Sborlini:2014mpa}.

In this article, we compute polarized splitting functions in the triple-collinear limit at next-to-leading order (NLO) in QCD. For the sake of simplicity, we consider only processes involving at least one photon. Quark-started splitting processes are constrained by helicity conservation. So, they turn out to be proportional to the unpolarized splitting functions, which were computed in a previous article \cite{Sborlini:2014mpa}. Explicitly, it is possible to write
\beqn
{\textbf{P}}_{q \to a_1 \ldots a_m}(s,s') &=& \omega_q \delta_{s \ s'} \, \la \hat{P}_{q \to a_1 \ldots a_m} \ra  \, ,
\eeqn
where $\omega_q$ is the number of fermionic degrees of freedom\footnote{This property is not obvious in the context of dimensional regularization (DREG). The main inconvenient arises from the extension of $\gamma^5$ to a $\DST$-dimensional space-time, which introduces some ambiguities in the treatment of fermion polarizations. In particular, some interactions can violate helicity-conservation as we described in Ref. \cite{Sborlini:2013jba}.}. For this reason, we only consider the polarized splitting functions associated with the processes $\gamma \to q \bar{q} \gamma$, $\gamma \to q \bar{q} g$ and $g \to q \bar{q} \gamma$.

The outline of the paper is the following. In Section \ref{sec:details} we describe the computational techniques applied to obtain the results. They are based in an extension of the Passarino-Veltman procedure at amplitude level, combined with inversion rules and transcendentality classification. After that, we present results for photon-started processes in Section \ref{sec:photon}. In this section we also include a brief discussion about the structure of these expressions, in order to complement the one exhibited in Ref. \cite{Sborlini:2014mpa}. Then we discuss the polarized splitting function for $g \to q \bar{q} \gamma$ and its corresponding NLO corrections, in Section \ref{sec:gluon}. Finally, we present the conclusions in Section \ref{sec:conclusion}.

%20140729_1645: REV OK!

%***********************************************************************************************************************************************%
\section{Collinear limits and polarized splitting functions}
\label{sec:details}
Before focusing into the details of the computation of polarized splitting functions, let's recall some useful definitions to analyse the multiple collinear limit. Let's consider an $n$-particle process where $m$ particles become collinear at the same time. Collinear momenta are labelled as $p_i$ with $i\in C=\{1,2, \ldots, m\}$ and these vectors fulfil $p_i^2=0$ (massless on-shell partons). The subenergies are defined as $s_{i j} = 2 \, p_i \cdot p_j$ and $s_{i,j} = \left(p_i +p_{i+1} + \ldots + p_{j} \right)^2 = p_{i,j}^2$. To avoid potential \textit{strict} factorization breaking issues \cite{Catani:2011st,Forshaw:2012bi}, we work in the time-like (TL) region, which implies $s_{ij} \geq 0$ for every $i,j \in C$. We mention strict factorization breaking effects in the context of the discussion presented in Ref. \cite{Catani:2011st}. In that article, the authors show that in space-like (SL) kinematics (i.e. $s_{ij} \geq 0$ and $s_{ik} \leq 0$ for some $i,j,k \in C$) some color correlations involving non-collinear partons can appear in the factorization formula. Also they have shown that the splitting amplitudes are independent of non-collinear particles only in the TL kinematics.

A proper description of the collinear limit requires the introduction of a pair of light-like vectors ($\wp^2 = 0$, $n^2 = 0$), such that 
\beq
\wp^{\mu} = p_{1,m}^{\mu} - \frac{s_{1,m}}{2 \; n \cdot \wp} \, n^{\mu} \, ,
\label{ptilde}
\eeq
corresponds to the collinear direction in the multiparton collinear limit, and $n^\mu$ parametrizes how this limit is approached, with $n\cdot\wp = n\cdot p_{1,m}$. The longitudinal-momentum fractions $z_i$ are given by
\beq
z_i = \frac{n \cdotp p_i}{n\cdot \wp}~, \qquad i \in C~,
\eeq
and they fulfil the constraint $\sum_{i\in C} z_i = 1$.

Factorization properties become explicit when virtual gluons or photons are allowed to have only physical polarizations. For this reason, we work in the light-cone gauge (LCG), which is characterized by the absence of ghosts and
\beq
d^{\mu \nu} (k,Q) = - \eta^{\mu \nu} + \frac{k^\mu Q^\nu + Q^\mu k^\nu}{Q\cdot k} \,
\label{bosonpol}
\eeq 
is the physical polarization tensor of a gauge vector boson (gluon or photon) with momentum $k$ and $Q^2=0$, $k \cdot Q \neq 0$. Although the quantization vector $Q$ is arbitrary, we choose $Q=n$ in order to simplify the computation.

Polarized splitting functions are obtained from the tensor product of two amputated splitting matrices. Using collinear factorization properties~\cite{Collins:1989gx, Catani:2011st}, we know that
\beqn
\ket{{\cal A}\left(p_1, \ldots, p_n\right)} &\simeq& \sp_{a \to a_1 \ldots a_m}(p_1 ,\ldots, p_m; \wp) \, \ket{{\cal A}(\wp, p_{m+1}, \ldots, p_n)}~,  
\label{FACTORIZACIONLOmultiple1}
\eeqn
where the sum over the physical polarizations of the intermediate parent parton is understood. The index $a$ is fixed by flavour conservation for processes started by QCD partons, so we drop this label for these configurations. Since we are considering also photon initiated processes, it has to be explicitly specified in our notation to avoid ambiguities. So, in that case we write $\gamma \to a_1 \ldots a_m$. In order to make a complete general analysis, we kept in this section the complete flavour labelling to treat simultaneously gluon and photon-started splitting processes. It is important to notice that \Eq{FACTORIZACIONLOmultiple1} only takes into account the most divergent contributions in the limit $s_{1,m} \to 0$, neglecting all the subleading terms. Besides that, it constitutes a definition of the splitting matrices $\sp$ in the color$+$spin space.
%%%%%%%%%%%%%%%%%%%%%%%%%%%%%%%%%%%%%%%%%%%%%%%%%%%%%%%%%%%

Relying on the previously mentioned collinear factorization properties, let's explain how to compute the spin-dependent splitting functions. With the aim of disentangling the different helicity contributions, we remove the polarization vector from the splitting amplitude. Explicitly,
\beqn
\ket{{\cal A}\left(p_1, \ldots, p_n\right)} &\simeq& \sum_{\lambda \in {\rm phys.pol.}}\sp^{\mu}_{a \to a_1 \ldots a_m}(p_1 ,\ldots, p_m; \wp) \epsilon_{\mu}(\wp,\lambda) \, \ket{{\cal A}(\wp^{-\lambda}, p_{m+1}, \ldots, p_n)} ~,  
\label{FACTORIZACIONLOmultiple2}
\eeqn
thus, after taking the square of this formula, we obtain
\beqn
\nn \bra{{\cal A}\left(p_1, \ldots, p_n\right)} &\IDC& \ket{{\cal A}\left(p_1, \ldots, p_n\right)} \simeq \sum_{\lambda,\lambda'} \bra{{\cal A}(\wp^{-\lambda}, p_{m+1}, \ldots, p_n)} 
\\ \nn &\times& \, (\epsilon_{\mu}(\wp,\lambda))^{*} \left(\sp^{\mu}_{a \to a_1 \ldots a_m}(p_1 ,\ldots, p_m; \wp)\right)^{\dagger} \sp^{\nu}_{a \to a_1 \ldots a_m}(p_1 ,\ldots, p_m; \wp) \epsilon_{\nu}(\wp,\lambda') \,
\\ &\times& \ket{{\cal A}(\wp^{-\lambda'}, p_{m+1}, \ldots, p_n)} ~,  
\label{FACTORIZACIONNLOmultipleSQ}
\eeqn
which allows to define the polarized splitting function according to
\beqn
P^{\mu \nu}_{a \to a_1 \ldots a_m} &\equiv& \left( \frac{s_{1,m}}{2 \;\mu^{2\ep}} \right)^{m-1} \; \left(\sp^{\mu}_{a\to a_1 \ldots a_m}(p_1 ,\ldots, p_m; \wp)\right)^{\dagger} \sp^{\nu}_{a\to a_1 \ldots a_m}(p_1 ,\ldots, p_m; \wp) \,
\label{PpolarizedLOdefinition1}
\eeqn
that represents the product of two amputated splitting matrices. This product implies a sum over polarizations (and colors) of all the outgoing collinear partons, but parent parton polarization is not specified. The presence of a mass scale $\mu$ in the normalization of the splitting functions is related with the fact that we use dimensional regularization (DREG) \cite{Bollini:1972ui, 'tHooft:1972fi} with $\DST=4-2\ep$ space-time dimensions. Also, it is crucial to appreciate that the collinear limit is completely described by the object 
\beqn
{\textbf{P}}_{a \to a_1 \ldots a_m}(\lambda,\lambda') &=& (\epsilon_{\mu}(\wp,\lambda))^{*}  P^{\mu \nu}_{a \to a_1 \ldots a_m} \epsilon_{\nu}(\wp,\lambda') \, ,
\label{PpolarizedLOdefinition2}
\eeqn
which implies that we drop terms proportional to $\wp^{\mu}$ and $n^{\mu}$ in the tensorial expansion of $P^{\mu \nu}_{a \to a_1 \ldots a_m}$, because $\epsilon^{\mu}(\wp,\lambda)$ is associated to an on-shell vector particle in a physical gauge and it must fulfil 
\beqn
\epsilon(\wp,\lambda) \cdot n = \epsilon(\wp,\lambda) \cdot \wp &=& 0 \,
\label{EcuacionRequisitosGAUGE}
\eeqn
for every physical polarization $\lambda$. In consequence, combining \Eq{ptilde} and \Eq{EcuacionRequisitosGAUGE}, we conclude that it is possible to make the replacement $p_m^{\mu} = - p_{1,m-1}^{\mu}$ and cancel terms proportional to $n^{\mu}$. We anticipate that this fact will allow us to reduce the size of the tensorial basis employed to expand the polarized splitting functions.

Since the computation of the collinear limit of squared amplitudes can be done using amputated amplitudes, then it is preferable to express our results in terms of $P_{a\to a_1 \ldots a_m}^{\mu \nu}$. Of course, in the helicity formalism, it is more suitable to consider ${\textbf{P}}_{a \to a_1 \ldots a_m}$. In any case, both expressions can be easily related by contracting with polarization vectors or just by removing them.

Considering collinear factorization at one-loop level,
\beqn
\nn \ket{{\cal A}^{(1)}\left(p_1, \ldots, p_n\right)} &\simeq& \sp^{(1)}_{a \to a_1 \ldots a_m}(p_1 ,\ldots, p_m; \wp) \, \ket{{\cal A}^{(0)}(\wp, p_{m+1}, \ldots, p_n)}
\\ &+& \sp^{(0)}_{a \to a_1 \ldots a_m}(p_1 ,\ldots, p_m; \wp) \, \ket{{\cal A}^{(1)}(\wp, p_{m+1}, \ldots, p_n)} ~,  
\label{FACTORIZACIONNLOmultiple}
\eeqn
then the one-loop correction to the polarized splitting function is given by
\beqn
\nn P^{(1), \mu \nu}_{a \to a_1 \ldots a_m} &\equiv& \left( \frac{s_{1,m}}{2 \;\mu^{2\ep}} \right)^{m-1} \; \left(\sp^{(0),\mu}_{a \to a_1 \ldots a_m}(p_1 ,\ldots, p_m; \wp)\right)^{\dagger} 
\\ &\times& \sp^{(1),\nu}_{a \to a_1 \ldots a_m}(p_1 ,\ldots, p_m; \wp) + \, {\rm h.c. }\, .
\label{PpolarizedNLOdefinition1}
\eeqn
We will use this expression as a master formula for all our calculations.

%%%%%%%%%%%%%%%%%%%%%%%%%%%%%%%%%%%%%%%%%%%%%%%%%%%%%%%%%%%%%%%%%%%%%%%%%%%%%%%%%%%%%%%%%%%%%%%%%%%%%%%%%%%%%%%%%%%%%%%%%%
%%%%%%%%%%%%%%%%%%%%%%%%%%%%%%%%%%%%%%%%%%%%%%%%%%%%%%%%%%%%%%%%%%%%%%%%%%%%%%%%%%%%%%%%%%%%%%%%%%%%%%%%%%%%%%%%%%%%%%%%%%

After introducing a general definition for $P_{a \to a_1 \ldots a_m}^{\mu \nu}$, a tensorial basis is required to perform an expansion of this object. When considering an $n$-particle process with $m$-collinear partons, there are $m$ vectors associated with external momenta and a null-vector $n^{\mu}$ introduced by the quantization procedure. Due to the fact that $P_{a \to a_1 \ldots a_m}^{\mu \nu}$ is a rank-$2$ tensor and it depends only on $\eta^{\mu \nu}$, $\left\{p_i^{\mu}\right\}_{i\in C}$ and $n^{\mu}$, then we define the basis
\beqn
f_{1}^{\mu\nu} &=& \eta_{\DST}^{\mu \nu} \, ,
\\ f_{1+i}^{\mu\nu} &=& \tilde{p}_{\sigma_1(i),\sigma_2(i)}^{\mu \nu} \ \ i \in \left\{1,\ldots,\Delta_1\right\} \, ,
\\ f_{1+i+\Delta_1}^{\mu\nu} &=& \bar{p}_{\rho_1(i),\rho_2(i)}^{\mu \nu} \ \ i \in \left\{1,\ldots,\Delta_2\right\} \, ,
\\ f_{1+j+\Delta_1+\Delta_2}^{\mu\nu} &=& \tilde{p}_{j,m+1}^{\mu \nu} \ \ \ \ \ \ j \in \left\{1,\ldots,m\right\} \, ,
\\ f_{1+j+\Delta_1+\Delta_2+m}^{\mu\nu} &=& \bar{p}_{j,m+1}^{\mu \nu} \ \ \ \ \ \ j \in \left\{1,\ldots,m\right\} \, ,
\\ f_{2+\Delta_1+\Delta_2+2m}^{\mu\nu} &=& \tilde{p}_{m+1,m+1}^{\mu \nu} \, ,
\eeqn
with
\beqn
\tilde{p}_{i,j}^{\mu\nu} &=& p_i^{\mu}p_j^{\nu}+p_j^{\mu}p_i^{\nu} \, ,
\\ \bar{p}_{i,j}^{\mu\nu} &=& p_i^{\mu}p_j^{\nu}-p_j^{\mu}p_i^{\nu} \, ,
\\ \Delta_1 &=& \frac{m(m+1)}{2} \, ,
\\ \Delta_2 &=& \frac{m(m-1)}{2} \, ,
\eeqn
where we define $p_{m+1}^{\mu}=n^{\mu}$ to simplify the notation.\footnote{The validity of this assumption is restricted to TL kinematics. Otherwise, factorization breaking effects described in Refs. \cite{Catani:2011st, Forshaw:2012bi} could introduce a dependence in the non-collinear partons.} In the previous expressions, $\sigma$ is a permutation of pairs of collinear momenta which can also include repeated elements and contributes to the symmetric part; $\rho$ is a permutation that excludes repeated indices. Also it is important to appreciate that $f_1$ is the $\DST$-dimensional metric tensor, with $\DST$ the number of space-time dimensions. When using DREG, we could choose $f^{\mu \nu}_1=\eta_4^{\mu\nu}$ and it would be associated with a different regularization scheme (RS). In the context of conventional dimensional regularization (CDR), it is requested to use $f^{\mu \nu}_1=\eta^{\mu \nu}_{\DST}$ to achieve consistency.

It is worth noticing that, in spite of imposing the cancellations induced by the contraction with $\epsilon^{\mu}(\wp)$, we can not completely neglect the remaining elements in the basis. The reason is that the computation of tensor-like integrals requires a \textit{complete} basis of tensorial structures. In other words, when performing the tensorial reduction it is mandatory to project over all the possible tensor products of the vectors involved in the integral and the metric tensor (whenever considering a rank higher than or equal to $2$). In consequence, working at the integrand level and using \Eq{EcuacionRequisitosGAUGE}, we can throw away \textit{a posteriori} contributions proportional to $n^{\mu}$ and $\wp^{\mu}$ when $\mu$ is an index referring to the parent parton's polarization vector. After expanding the general expression for $P^{\mu \nu}_{a \to a_1 \ldots a_m}$ and applying this procedure, we obtain
\beqn
\nn P^{\mu \nu}_{a \to a_1 \ldots a_m} &=& \sum_{j=1}^{1+\Delta_1+\Delta_2} \left(\int_q A^{(0)}(q)\right) f_j^{\mu \nu} \left.\right|_{{\cal S}_{\mu} \cup {\cal S}_{\nu}} + \sum_{j=1}^{m} \left(\int_q A^{(1)}(q) q^{\nu} \right) p_j^{\mu} \left.\right|_{{\cal S}_{\mu}}
\\ &+& \sum_{j=1}^{m} \left(\int_q A^{(2)}(q) q^{\mu} \right) p_j^{\nu} \left.\right|_{{\cal S}_{\nu}} \,\, + \int_q A^{(3)}(q) q^{\mu}q^{\nu} \, , 
\label{Descomposition1}
\eeqn
where we defined
\beqn
\int_q &=& - \imath \, \int \, \frac{d^{\DST}q}{(2\pi)^{\DST}} \, ,
\eeqn
and $A^{(l)}(q)$ is a scalar function of the loop momenta. In \Eq{Descomposition1}, ${{\cal S}_{\sigma}}$ implements all the cancellations associated to index $\sigma=\{\mu,\nu\}$, i.e.
\beqn
\left. n^{\sigma} \right|_{{\cal S}_{\sigma}} &\to& 0 \, ,
\\ \left. \wp^{\sigma} \right|_{{\cal S}_{\sigma}} &\to& 0 \, ,
\\ \left. p_{1,m}^{\sigma} \right|_{{\cal S}_{\sigma}} &\to& 0 \, ,
\\ \left. p_{m}^{\sigma} \right|_{{\cal S}_{\sigma}} &\to& - \sum_{i=1}^{m-1} p_{i}^{\sigma} \, ,
\eeqn
which are consequence of \Eq{EcuacionRequisitosGAUGE} and the definition of $\wp$. The first term in the r.h.s of \Eq{Descomposition1} contains only scalar integrals and the tensorial dependence is independent of the loop momentum. Then, we neglect those elements of the basis which contains $n^{\sigma}$ because they will be cancelled after applying ${\cal S}_{\mu} \cup {\cal S}_{\nu}$. In consequence, it is enough to sum over the first $1+\Delta_1+\Delta_2$ elements of the basis. In a similar way, the second and third terms of \Eq{Descomposition1} involve rank-1 tensor integrals and we can exclude $n^{\sigma}$ in the sums. The last term contains rank-2 integrals and the cancellations must be imposed \textit{after} performing the tensorial reduction. In summary, \Eq{Descomposition1} describes all the simplifications that can be carried out before applying any tensorial-reduction technique, decreasing the complexity of the intermediate steps of the computation.

We would like to emphasize that our approach is different from the usual Passarino-Veltman reduction, since we are not treating integrals as isolated objects. Instead, we are combining them inside the scattering amplitude and, then performing the reduction simultaneously. This method is more efficient because it exploits the symmetries associated with the matrix elements. In both cases, it is mandatory to employ a complete basis to write tensor integrals.

The following step consists in projecting \Eq{Descomposition1} over the $\upsilon=((m+1)^2+1)$ elements of the whole basis. So, we get
\beqn
P^{\mu \nu}_{a \to a_1 \ldots a_m} &=& \sum_{j=1}^{\upsilon} \, A_j f_j^{\mu \nu} \, ,
\label{Descomposition2}
\eeqn
and we define the vector $B_j$ as
\beqn
B_j &=& \sum_{i=1}^{\upsilon} \, A_i f_i^{\mu \nu} {(f_j)}_{\mu\nu} = \left(M \cdot A\right)_j \, ,
\label{VectorB}
\eeqn
with the kinematic matrix $(M)_{ij} = f_i^{\mu \nu} {(f_j)}_{\mu\nu}$. It is important to note that this $\upsilon\time\upsilon$-dimensional matrix contains information about all the possible scalar products among collinear particle momenta and $n^\sigma$, together with $\left(\eta_{\DST}\right)^{\mu}_{\mu}=\DST$ (the trace of the $\DST$-dimensional metric tensor). Also, if $\DST=4$ this matrix becomes singular because momenta are not represented by independent vectors. For this reason, ${\rm Det}(M)=\cal{O}(\epsilon)$ when $\DST=4-2\epsilon$. Of course, through the computation of $M^{-1}$ we recover the coefficients in the expansion \Eq{Descomposition2} but this procedure is extremely lengthy due to the size of $M$.

In the special case of the triple collinear limit, we use Cramer's rule to recover the coefficients inside \Eq{Descomposition2}. First of all, we rewrite the tensorial basis making a distinction according to the symmetry properties. Thus
\beqn
f_{1}^{\mu\nu} &=& \eta_{\DST}^{\mu\nu} \, , 
\\ f_{2}^{\mu\nu} &=& \frac{p_1^{\mu}p_2^{\nu}+p_1^{\nu}p_2^{\mu}}{s_{123}} \, , 
\\ f_{3}^{\mu\nu} &=& 2 \frac{p_1^{\mu}p_1^{\nu}}{s_{123}} \, , 
\\ f_{4}^{\mu\nu} &=& 2 \frac{p_2^{\mu}p_2^{\nu}}{s_{123}} \, , 
\\ f_{5}^{\mu\nu} &=& \frac{p_1^{\mu}p_{123}^{\nu}+p_1^{\nu}p_{123}^{\mu}}{s_{123}} \, ,
\\ f_{6}^{\mu\nu} &=& \frac{p_2^{\mu}p_{123}^{\nu}+p_2^{\nu}p_{123}^{\mu}}{s_{123}} \, ,
\\ f_{7}^{\mu\nu} &=& 2\frac{p_{123}^{\mu}p_{123}^{\nu}}{s_{123}} \, ,
\\ f_{8}^{\mu\nu} &=& \frac{p_1^{\mu}n^{\nu}+p_1^{\nu}n^{\mu}}{n\cdot \wp} \, ,
\\ f_{9}^{\mu\nu} &=& \frac{p_2^{\mu}n^{\nu}+p_2^{\nu}n^{\mu}}{n\cdot \wp} \, ,
\\ f_{10}^{\mu\nu} &=& \frac{p_{123}^{\mu}n^{\nu}+p_{123}^{\nu}n^{\mu}}{n\cdot \wp} \, ,
\\ f_{11}^{\mu\nu} &=& s_{123}\, \frac{n^{\mu}n^{\nu}}{(n\cdot \wp)^2} \, ,
\eeqn
are the symmetric structures, while
\beqn
f_{12}^{\mu\nu} &=& \frac{p_1^{\mu}p_2^{\nu}-p_1^{\nu}p_2^{\mu}}{s_{123}} \, ,
\\ f_{13}^{\mu\nu} &=& \frac{p_1^{\mu}p_{123}^{\nu}-p_1^{\nu}p_{123}^{\mu}}{s_{123}} \, ,
\\ f_{14}^{\mu\nu} &=& \frac{p_2^{\mu}p_{123}^{\nu}-p_2^{\nu}p_{123}^{\mu}}{s_{123}} \, ,
\\ f_{15}^{\mu\nu} &=& \frac{p_1^{\mu}n^{\nu}-p_1^{\nu}n^{\mu}}{n\cdot \wp} \, ,
\\ f_{16}^{\mu\nu} &=& \frac{p_2^{\mu}n^{\nu}-p_2^{\nu}n^{\mu}}{n\cdot \wp} \, ,
\\ f_{17}^{\mu\nu} &=& \frac{p_{123}^{\mu}n^{\nu}-p_{123}^{\nu}n^{\mu}}{n\cdot \wp} \, ,
\eeqn
give rise to the antisymmetric ones. Notice that all the basis elements are dimensionless quantities. Since symmetric and antisymmetric spaces are orthogonal, the matrix $M$ in \Eq{VectorB} can be written as 
\beqn
M &=& \left(
\begin{array}{cc}
 M_{\rm sym} & 0 \\
 0 & M_{\rm asym} \\
\end{array}
\right) \, ,
\eeqn
where $M_{\rm sym}$ is a $11\times 11$ matrix while $M_{\rm asym}$ has dimension $6 \times 6$. We are going to treat both contributions independently.

As mentioned before, the determinant of $M$ vanishes in the limit $\epsilon \to 0$. Explicitly,
\beqn
\det (M) &=& \det \left(M_{\rm sym}\right) \times \det \left(M_{\rm asym}\right) \, ,
\\ \det \left(M_{\rm asym}\right) &=& \Omega^3  \, ,
\\ \det \left(M_{\rm sym}\right) &=& - 8 \epsilon \, \Omega^5   \, ,
\eeqn
and
\beqn
\Omega &=& \sum_{i=1}^3 x_i z_i \left(x_i z_i - \sum_{j \neq i} x_j z_j\right) \, ,
\eeqn
with the notation
\beqn
x_{i} &=& \frac{-s_{j k}- \imath 0}{-s_{123}- \imath 0}~,
\eeqn
where $(i,j,k)$ is a reordering of the indices set $\left\{1,2,3\right\}$. $\Omega$ is independent of $\epsilon$ and cyclically invariant under relabelling of particles. Also, it is important to appreciate that $M$ becomes singular when working in $\DST=4$ due to the linear dependence on the momenta.

After specifying the tensor basis, we introduce the vector $B_j$ following \Eq{VectorB}. Due to the cancellations mentioned before, we just need to know $4$ coefficients for the symmetric part and only $1$ for the antisymmetric one. In other terms, we can expand the polarized splitting function as
\beqn
P^{\mu \nu}_{a \to a_1 a_2 a_3} &=& \sum_{j=1}^{4} \, A^{\rm sym}_j f_j^{\mu \nu} \, + \, A^{\rm asym} f_{12}^{\mu \nu} \, ,
\label{ExpansionPmunuTRIPLE1}
\eeqn
after neglecting contributions that are proportional to $n^{\mu}$ and $p^{\mu}_{123}$. To obtain the coefficients $A_j^{\rm sym}$ and $A^{\rm asym}$ we use Cramer's rule by introducing the matrices
\beqn
\left(M^{\rm Cramer}_{\rm sym}\right)_{ij} &=& -\frac{\det{\bar{M}^{(i,j)}}}{8 \epsilon \Omega^8} \ \ \ \ \ \ i \in \left\{1,\ldots,4\right\} \ , \ j \in \left\{1,\ldots,17\right\} \,,
\\ \left(M^{\rm Cramer}_{\rm asym}\right)_{j} &=& -\frac{\det{\bar{M}^{(12,j)}}}{8 \epsilon \Omega^8} \ \ \ \ \ \ \ \ j \in \left\{1,\ldots,17\right\}\,,
\eeqn
where $\bar{M}^{(i,j)}$ denotes a new matrix formed by replacing the column $i$ of $M$ with the canonical vector $\hat{e}_j$. Thus, $M^{\rm Cramer}_{\rm sym}$ is a $4 \times 17$-dimensional matrix while $M^{\rm Cramer}_{\rm asym}$ is just a $17$-dimensional vector. These matrices allow us to recover only the relevant coefficients, which makes this approach more efficient than inverting the whole system. So,
\beqn
A^{\rm sym}_j &=& \left(M^{\rm Cramer}_{\rm sym} \cdot B\right)_{j} \ \ \ \ j \in \left\{1,\ldots,4\right\} \,, 
\\ A^{\rm asym} &=& \left(M^{\rm Cramer}_{\rm asym} \cdot B\right) \, ,
\eeqn
lead to the desired expressions.

Finally we would like to make some remarks about the treatment of $B_j$. Since each component of this vector is a scalar, all this procedure simplifies the computation of Feynman integrals due to the presence of only scalar ones. We must take into account the existence of different propagators contributing to $B_j$. For this reason, we define certain propagator's basis and we put together all the contributions that can be described inside the same set. Then, integration by parts (IBP) reduction \cite{Chetyrkin:1981qh, Laporta:2001dd} is applied and all the components are expanded using a set of master integrals.

%20140728_1835: Casi terminado
%20140730_1152: OK!!!
On the other hand, $\sp_{a \to a_1 \ldots a_m}^{(1)}$ can be decomposed as
\beqn
\sp^{(1)}_{a \to a_1 \dots a_m} &=& \sp^{(1)\,{\rm div.}}_{a \to a_1 \dots a_m} + \sp^{(1)\,{\rm fin.}}_{a \to a_1 \dots a_m}~,
\label{DescomposicionSP1}
\eeqn
where $\sp^{(1)\,{\rm fin.}}_{a \to a_1 \dots a_m}$ contains only finite pieces while IR/UV divergences are kept inside $\sp^{(1)\,{\rm div.}}_{a \to a_1 \dots a_m} $. Moreover, $\sp^{(1)\,{\rm div.}}_{a \to a_1 \dots a_m}$ can be expressed as
\beq 
\sp^{(1)\,{\rm div.}}_{a \to a_1 \dots a_m} (p_1,\dots,p_m; \wp) =
\IC^{(1)}_{a \to a_1 \dots a_m}(p_1,\dots,p_m;\wp) \, \sp^{(0)}_{a \to a_1 \dots a_m}(p_1, \dots, p_m; \wp)~,
\label{DescomposicionSP2}
\eeq
with the insertion operator
\beeq 
\IC^{(1)}_{a \to a_1 \dots a_m}(p_1,\dots,p_m;\wp) &=& \CG \, \g^2 \, \left( \frac{-s_{1, m} -i0}{\mu^2}\right)^{-\ep} \, 
\Bigg\{ \frac{1}{\ep^2}
\sum_{i,j=1 (i \neq j)}^{\bar{m}} \;{\bom T}_i \cdot {\bom T}_j
\left( \frac{-s_{ij} -i0}{-s_{1, m} -i0}\right)^{-\ep} \nn \\
&+&
\frac{1}{\ep^2}
\sum_{i,j=1}^{\bar{m}} \;{\bom T}_i \cdot {\bom T}_j
\;\left( 2 - \left( z_i \right)^{-\ep} -\left( z_j \right)^{-\ep} \right) \nn \\
&-& \frac{1}{\ep} 
\left( \sum_{i=1}^{\bar{m}} \left( \gamma_i - \ep {\tilde \gamma}_i^{\RS} \right)
- \left( \gamma_a - \ep {\tilde \gamma}_a^{\RS} 
\right) - \frac{\tilde{m}-2}{2} \left( \beta_0 - \ep {\tilde \beta}_0^{\RS}
\right) \right) \Bigg\}~,
\label{sp1div}
\eeeq
where the color matrix of the collinear particle with momentum $p_i$ is denoted by ${\bom T}_i$, $\bar{m}$ counts the number of collinear final state QCD partons and $\tilde{m}$ refers to the total number of QCD partons in the splitting process. This formula was first introduced in Ref. \cite{Catani:2003vu} and constitutes an extension to the multiple collinear limit of the original one derived by Catani and Seymour\footnote{For more details, see Ref. \cite{Catani:1996vz}, Section $7.3$.}. However, there are some alternative approaches that lead to similar expressions. For instance in Ref. \cite{Becher:2009qa}, the authors use renormalization techniques for the treatment of IR singularities; proposing an all-order formula for the anomalous dimension, they obtained a general structure for the IR-divergences of scattering amplitudes.

The description of color operators is based on the discussion presented in Refs. \cite{Catani:1996vz, Catani:1998bh}. In the context of QCD with $N_{\rm C}$ colors, the associated gauge group is ${\rm SU}(N_{\rm C})$. Given a representation $R$ of the algebra, the generators are normalized according to
\beqn
{\rm Tr}\left[\SUNT^a(R) \SUNT^b(R)\right] &=& T_R \, \delta^{ab} \, .
\eeqn
As a conventional choice, we use $T_A=N_{\rm C}$ and $T_F=1/2$ for the adjoint and fundamental representations, respectively. Applying Fierz identities and the definition of Casimir operators implies
\beqn
C_A &=& N_{\rm C} \, ,
\\ C_F &=& \frac{N^2_{\rm C}-1}{2 N_{\rm C}} \, ,
\eeqn
and also we use ${\rm Tr}\left[\IDC\right]=C_A$, with $\IDC$ the identity element in the fundamental representation.

Following with the description of \Eq{sp1div}, the one-loop $\DST$-dimensional volume factor is given by
\beq
\CG = \frac{\Gamma\left(1+\ep\right)\, \Gamma\left(1-\ep\right)^2}{\left(4 \pi\right)^{2-\ep}\, \Gamma\left(1-2\ep\right)} \, ,
\eeq
and final state particles are ordered such that $\left\{1, \ldots, \bar{m}\right\}$ are the coloured ones while the remaining ones are singlets under ${\rm SU}(N_{\rm C})$ transformations. Also, it is useful to notice that $\tilde{m}=\bar{m}$ in the collinear splitting processes which are started by non-QCD partons (in this paper, photons). On the other hand, the flavour coefficients $\gamma_a$ are given by
\beqn
\gamma_q=\gamma_{\bar q}&=&3C_F/2 \, ,
\\ \gamma_g &=& \beta_0/2 \, ,
\eeqn
and $\beta_0=(11C_A-2N_f)/3$, while $\gamma_a=0$ for non-QCD partons. Besides predicting the $\epsilon$-poles, $\IC^{(1)}$ also controls the RS dependence up to ${\cal O}(\ep^0)$ through the coefficients ${\tilde \gamma}_i^{\RS}$ and ${\tilde \beta}_0^{\RS}$. They are given by
\beq
{\tilde \gamma}_i^{\cdr}={\tilde \beta}_0^{\cdr}=0 \; ,
\eeq
in CDR, while
\beqn
{\tilde \gamma}_q^{\dimr}&=&{\tilde \gamma}_{\bar q}^{\dimr}=C_F/2 \; , 
\\ {\tilde \gamma}_g^{\dimr}&=&{\tilde \beta}_0^{\dimr}/2=C_A/6 \; ,
\eeqn
in dimensional reduction (DR).

As can be seen from \Eq{sp1div}, all the divergent structure is controlled by the insertion operator $\IC^{(1)}_{a \to a_1 \dots a_m}$. This object is a matrix in the color space, but for the processes considered it is possible to completely describe its action using a pure $c$-number. Let's explain this point more carefully. Due to color conservation, we have
\beq
\sum_i {\bom T}_i \;\sp^{(0)}_{a \to a_1 \dots a_m} = \sp^{(0)}_{a \to a_1 \dots a_m} \;{\bom T}_a~,
\label{ColorCHARGEparent}
\eeq
thus the color charge of the parent parton can be expressed using the color information of the outgoing collinear particles. When $\tilde{m} \leq 3$, \Eq{ColorCHARGEparent} implies that all the products of color operators inside $\IC^{(1)}_{a \to a_1 \dots a_m} $ are proportional to the unit matrix. So, we write
\beq
\IC^{(1)}_{a \to a_1 \dots a_m} \to I^{(1)}_{a \to a_1 \dots a_m} \IDC \, ,
\eeq
where $I^{(1)}_{a \to a_1 \dots a_m}$ is a pure $c$-number.

After discussing the divergent structure of splitting functions at NLO, we can exploit this knowledge to write the finite corrections in an advantageous way. If we apply the decomposition suggested in \Eq{DescomposicionSP1} and \Eq{DescomposicionSP2} to the definition given in \Eq{PpolarizedNLOdefinition1}, we obtain
\beqn
\nn P^{(1), \mu \nu}_{a \to a_1 \ldots a_m} &\equiv& \left( \frac{s_{1,m}}{2 \;\mu^{2\ep}} \right)^{m-1} \; \left(\sp^{(0),\mu}_{a \to a_1 \dots a_m}\right)^{\dagger} \left( \sp^{(1)\,{\rm div.},\nu}_{a \to a_1 \dots a_m} + \sp^{(1)\,{\rm fin.},\nu}_{a \to a_1 \dots a_m} \right) + \, {\rm h.c. }\, ,
\\ &=& 2 \, {\rm Re}\left( I^{(1)}_{a \to a_1 \cdots a_m}(p_1,\dots,p_m;\wp) \right) 
 P^{(0), \mu \nu}_{a \to a_1 \ldots a_m} + \left( P^{(1)\,{\rm fin.},\mu \nu}_{a \to a_1 \ldots a_m} +  {\rm c.c.}\right)~,
\label{PpolarizedNLOdefinitionRENORMALIZED}
\eeqn
with
\beqn
P^{(1)\,{\rm fin.},\mu \nu}_{a \to a_1 \ldots a_m} &=& \left( \frac{s_{1,m}}{2 \;\mu^{2\ep}} \right)^{m-1} \; \left(\sp^{(0),\mu}_{a \to a_1 \dots a_m}\right)^{\dagger} \sp^{(1)\,{\rm fin.},\nu}_{a \to a_1 \dots a_m} \, ,
\eeqn
where we must recall that a sum over color and polarization of outgoing collinear particles is always understood. Centering in the triple collinear limit, \Eq{ExpansionPmunuTRIPLE1} can be rewritten as
\beqn
P^{(1)\,{\rm fin.},\mu \nu}_{a \to a_1 a_2 a_3} &=& c^{a \to a_1 a_2 a_3} \left[ \sum_{j=1}^{4} \, A^{(1)\,{\rm fin.}}_j f_{j}^{\mu \nu} \, + \, A^{(1)\,{\rm fin.}}_5 f_{12}^{\mu \nu} \right] \, ,
\label{EquacionDescomposicion2}
\eeqn
with $c^{a \to a_1 \cdots a_m}$ is a normalization factor which depends on the process and, at this point, we can just take care of the coefficients $A_i$. Since all the processes studied in this work are of the form $V \to q_1 \bar{q}_2 V_3$, they turn out to be symmetric under the exchange $1 \leftrightarrow 2$. The tensorial basis has a well-defined behaviour under the symmetry operator $S_{1\leftrightarrow 2}$, closely related with the symmetry properties in the indices $\mu\leftrightarrow\nu$; explicitly,
\beqn
S_{1\leftrightarrow 2} \left(f^{\mu \nu}_{j}\right) &=& f^{\mu \nu}_{j} \ \ \ {\rm for} \ \ \ j\in\left\{1,2\right\} \, ,
\\ S_{1\leftrightarrow 2} \left(f^{\mu \nu}_{3}\right) &=& f^{\mu \nu}_{4}  \, ,
\\ S_{1\leftrightarrow 2} \left(f^{\mu \nu}_{12}\right) &=& - f^{\mu \nu}_{12}  \, ,
\eeqn
so we can infer the behaviour of the associated coefficients. Thus, $A^{(1)\,{\rm fin.}}_4$ is obtained from $A^{(1)\,{\rm fin.}}_3$. Of course, making no assumptions about the symmetry during the computation allows to check for potential errors at the final stage.

The last step in the organization of the finite pieces consists in classifying the different terms according to their transcendental weight. The notion of transcendental weight is related to the number of iterated integrals of rational functions required to express a specific function. In this way, rational functions (including constants) have weight $0$. $\log(x)$ and $\pi$ have weight $1$; ${\rm Li}_{n}(x)$ and $\zeta_n$ have weight $n$. Since it is a multiplicative quantity, $\log(x)\log(y)$ has weight $2$ and so on. It is known that one-loop QCD amplitudes can be expanded using up to weight $2$ functions, when considering only ${\cal O}\left(\epsilon^0\right)$ terms. For these reasons and symmetry considerations, the coefficients $A_j$ can be written as
\beqn
A^{(1)\,{\rm fin.}}_j &=& \sum_{i=0}^2 {\cal C}_j^{(i)}  \, + (1 \leftrightarrow 2) \,    \ {\rm for} \ j\in\left\{1,2\right\} \, ,
\\ A^{(1)\,{\rm fin.}}_3 &=& \sum_{i=0}^2 {\cal C}_3^{(i)}  \, ,
\\ A^{(1)\,{\rm fin.}}_5 &=& \sum_{i=0}^2 {\cal C}_5^{(i)}  \, - (1 \leftrightarrow 2)  \, ,
\eeqn
where ${\cal C}_j^{(i)}$ includes only functions of transcendental weight $i$.

As a final comment, let's notice that unpolarized splitting functions can be recovered by contracting $P^{\mu\nu}_{a \to a_1 \dots a_m} $ with $d_{\mu \nu}(\wp,n)$, i.e.
\beqn
\la \hat{P}_{a \to a_1 \cdots a_m} \ra &=& \frac{1}{\omega} \, d_{\mu \nu}(\wp,n) \,  P^{\mu \nu}_{a \to a_1 \ldots a_m} \, \, , 
\eeqn
where $\omega=2(1-\epsilon)$ is the number of physical polarizations associated with the parent vector particle.

%20140730_1602: OK!!!

%***********************************************************************************************************************************************%
\section{Photon-started processes}
%***********************************************************************************************************************************************%
%***********************************************************************************************************************************************%
\label{sec:photon}
In this section we present the results associated to the processes $\gamma \to q \bar{q} \gamma$ and $\gamma \to q \bar{q} g$. In contrast to the path followed in Ref. \cite{Sborlini:2014mpa}, we start analysing the simplest splitting process with the objective of improving our understanding of their structure.

%***********************************************************************************************************************************************%
\subsection{$\gamma \to q \bar{q} \gamma$}
\label{sec:AqqbarA}
%***********************************************************************************************************************************************%
Let's start with the $\gamma \to q \bar{q} \gamma$ splitting amplitude. It is the easiest process in the triple-collinear limit as it involves only Abelian interactions. At LO the splitting amplitude reads
\beqn
\Spamplitude^{(0)(a_1,a_2)}_{\gamma \to q_1 \bar{q}_2 \gamma_3} &=& \frac{e^2_q g^2_e \mu^{2\epsilon} \IDC_{a_1 a_2}}{ {s_{123}}} \, \bar{u}(p_1) 
\left( \frac{\slashed{\epsilon}(p_3) \slashed{p}_{13} \slashed{\epsilon}(\wp)}{s_{13}}-\frac{\slashed{\epsilon}(\wp) \slashed{p}_{23} \slashed{\epsilon}(p_3)}{s_{23}} \right) 
  \, v(p_2) \, \,
\, ,
\label{SplittingLOA-qqbA}
\eeqn
which implies that the LO polarized splitting function can be expressed as
\beqn
P^{(0),\mu \nu}_{\gamma \to q_1 \bar{q}_2 \gamma_3} &=& e_q^4 g_e^4 C_A \,  {\cal P}^{\mu\nu}\left(p_1,p_2,p_3;\wp\right) \, ,
\label{POLKERNELLOA-qqbAbis}
\eeqn
where we introduced the function
\beqn
\nn {\cal P}^{\mu\nu}\left(p_1,p_2,p_3;\wp\right) &=& \frac{1}{x_1 x_2} \left(\eta_{\DST}^{\mu\nu} \left(\epsilon x_1 (1-x_3)-(1-x_1)^2\right)+2\epsilon \, f_2^{\mu \nu}+2(\epsilon-1) \, f_3^{\mu \nu}\right)
\\ &+& (1 \leftrightarrow 2) \, .
\eeqn
Note that this expression is totally symmetric under the exchange of particles $1 \leftrightarrow 2$, and that it only involves symmetric elements of the tensorial basis. The function ${\cal P}$ describes completely the kinematics of all the splitting processes considered in this article. This is due to the factorization of the color structure at tree-level in the triple collinear limit with photons.

In spite of involving solely symmetric tensorial structures, NLO corrections include non-trivial contributions to ${\cal C}_5^{(i)}$. However, as expected, the full splitting function is completely symmetric under $1\leftrightarrow 2$. For $\gamma \to q \bar{q} \gamma$, the normalization factor is given by
\beqn
c^{\gamma \to q \bar{q} \gamma}&=& C_A C_F\, e_q^4 g_e^4 \g^2 \, ,
\label{NormalizacionAtoqqbarA}
\eeqn
while
\beqn
I^{(1)}_{\gamma \to q_1 \bar{q}_2 \gamma_3} (p_1, p_2, p_3; \wp) &=& \frac{\CG \g^2}{\ep^2} \left(\frac{-s_{123}-\imath 0}{\mu^2}\right)^{-\ep } \left[-2 C_F x_3^{-\ep} - 2 \ep \gamma_q\right] \, 
\label{eq:ICqqbargamma}
\eeqn
controls the divergent structure for this process. When comparing $\ep$-poles in our bare results with the ones predicted by this formula, we found a complete agreement.

Now, let's show the NLO corrections. The rational terms are described by
\beqn
{\cal C}_{1}^{(0)} &=& \frac{1-x_1}{x_1} \left(
\frac{8 (1-x_1)}{x_2} + 1 \right)~, \\ 
{\cal C}_{2}^{(0)} &=& \frac{4}{1-x_3} \left( \frac{1-x_1}{x_1\, x_2} - 1 \right)
- \frac{2}{1-x_1}~, \\ 
{\cal C}_{3}^{(0)} &=& \frac{1}{x_1\, x_2} \left( 
\frac{4 (1-x_2+x_2^2)}{1-x_3} 
+ \frac{(1-x_2)^2}{1-x_1} + 15 - x_2 \right)~, \\
{\cal C}_{5}^{(0)} &=& - \frac{2}{x_1} 
\left( \frac{1}{1-x_1} -\frac{2}{1-x_3} \right)~,
\eeqn
while
\beqn
{\cal C}_{1}^{(1)} &=& \frac{1-x_2}{x_2}
\left( \frac{2x_3-x_2}{1-x_1} \log (x_1) 
+ \frac{2 x_3}{1-x_3}  \log (x_3) \right)~, \\ 
\nn {\cal C}_{2}^{(1)} &=& 
\frac{2}{x_1\, x_2} \Bigg[ \frac{1}{1-x_1} \left( \frac{2x_3}{x_2}
-\frac{2x_1\, x_2 + x_3}{1-x_1}\right) \log (x_1) \nn \\
&+&\frac{2}{1 - x_3} \left( \frac{x_3}{x_1} 
+ \frac{x_3 (1 - x_1) - x_1 x_2}{1 - x_3} \right) \log (x_3) \Bigg]~,
\\ \nn {\cal C}_{3}^{(1)} &=& \frac{(1 - x_2)^2}{x_1\, x_2 (1 - x_1)} 
\left( \frac{1}{1 - x_1} + \frac{2}{x_2} \right) \log(x_1) 
+ \frac{2 x_3 - x_1}{x_1^2\, x_2} \log(x_2) \nn \\
&+& \frac{2}{(1 - x_3)^2} \left( 
\frac{2 (2 - x_2) x_3}{x_1\, x_2} + \frac{x_3^2}{x_1^2} 
+ \frac{1}{x_2^2} -  2 \right) \log(x_3)~, \\
{\cal C}_{5}^{(1)} &=&  \frac{2}{1-x_1} 
\left( \frac{2x_3}{x_1\, x_2} - \frac{1}{1-x_1} \right) \log (x_1)
+ \frac{4}{x_1 (1-x_3)^2}  \log (x_3)~,
\eeqn
contain the weight $1$ functions.  

Finally, for the weight $2$ contributions we have
\beqn
{\cal C}_{1}^{(2)} &=& \frac{2}{x_1\, x_2}
\left[(1-x_3) x_3 \left(1-\frac{1}{x_2}\right)-(1-x_1)^2\right] \, 
{\cal R}\left(x_1,x_3\right)~, \\
{\cal C}_{2}^{(2)} &=& \frac{4}{x_2^3}\left(1-\frac{(1-x_2)^2}{x_1}\right) \, 
{\cal R}\left(x_1,x_3\right)~, \\ 
{\cal C}_{3}^{(2)} &=& - \frac{2}{x_1 x_2}
\left[\left(2+\frac{(1-x_2)^2}{x_2^2}\right) {\cal R}\left(x_1,x_3\right)
+ \left(1+\frac{x_3^2}{x_1^2}\right) {\cal R}\left(x_2,x_3\right)
\right]~, \\ 
{\cal C}_{5}^{(2)} &=& -\frac{4 x_3}{x_1 \, x_2^2} \, 
{\cal R}\left(x_1,x_3\right)~,
\eeqn
with
\beqn
{{\cal R}\left(x_i,x_j\right)} &=& \frac{\pi^2}{6}-\ln{x_i}\ln{x_j}-\li{1-x_i}-\li{1-x_j}  \, ,
\label{DefinicionFUNCIONR}
\eeqn
being originated from the $\ep$-expansion of standard scalar boxes, after the subtraction of the terms proportional to $\log^2(x_i)$ included in $I^{(1)}_{\gamma \to q_1 \bar{q}_2 \gamma_3}(p_1, p_2, p_3; \wp)$. Explicitly, the standard scalar box with one off-shell leg is given by
\beqn
\nn I^{({\rm box})}_{ij} &=& \int_q \, \frac{\mu^{2\ep} s_{123}^2}{\left(q^2+\imath 0\right) \left((q-p_i)^2+\imath 0\right) \left((q-p_{123})^2+\imath 0\right) \left((q-p_{123}+p_j)^2 +\imath 0\right) } 
\\ \nn &=& \frac{2\CG}{\ep^2 \, x_i x_j} \left(\frac{-s_{123}-\imath 0}{\mu^{2}}\right)^{-\ep} \left[x_i^{-\ep} \, _2 F_1 \left(1,-\ep;1-\ep;-\frac{x_k}{x_j}\right)  + x_j^{-\ep} \, _2 F_1 \left(1,-\ep;1-\ep;-\frac{x_k}{x_i}\right) \right.
\\ &-& \left. \, _2 F_1 \left(1,-\ep;1-\ep;-\frac{x_k}{x_i x_j}\right)\right] \, ,
\label{DEFBoxes}
\eeqn
where $\{i,j,k\}$ is a permutation of $\{1,2,3\}$ and standard bubbles are simply written as
\beqn
I^{({\rm bubble})}_i &=& \int_q \, \frac{\mu^{2\ep}}{\left((q-p_i)^2 +\imath 0\right)\left((q-p_{123})^2+\imath 0\right)} = \frac{\CG}{\ep (1-2\ep)} \left(\frac{-s_{123}-\imath 0}{\mu^{2}}\right)^{-\ep} \, x_i^{-\ep} \, ,
\label{DEFBubbles}
\eeqn
with the notation $p_0^{\mu}=0^{\mu}$. Thus, we obtain the following identity 
\beqn
{{\cal R}\left(x_i,x_j\right)} &\equiv& - \frac{x_i x_j}{2} I^{({\rm box})}_{ij} + \frac{1-2\epsilon}{\epsilon} \left(I^{({\rm bubble})}_i+I^{({\rm bubble})}_j -I^{({\rm bubble})}_0\right) \, ,
\eeqn
which is valid up to ${\cal O}\left(\epsilon^0\right)$. This is an important step towards the extension of the results to higher orders in $\epsilon$, although rational coefficients dependence on $\epsilon$ could make it a bit complicated.

%20140730_1722: OK!!! Etapa1!
%***********************************************************************************************************************************************%
%***********************************************************************************************************************************************%
\subsection{$\gamma \to q \bar{q} g$}
\label{sec:Aqqbarg}
%***********************************************************************************************************************************************%
%***********************************************************************************************************************************************%
The following process is $\gamma \to q \bar{q} g$, which includes three QCD partons. Since all of them are on-shell final state particles, it is expected that the associated splitting function will be expressed in a very compact form. The corresponding splitting amplitude at tree-level is given by
\beqn
\nn \Spamplitude^{(0)(a_1,a_2,\alpha_3)}_{\gamma \to q_1 \bar{q}_2 g_3} &=& \frac{e_q g_e \g \mu^{2\ep} \SUNT^{\alpha_3}_{a_1 a_2}}{ {s_{123}}} \, \bar{u}(p_1) 
\left( \frac{\slashed{\ep}(p_3) \slashed{p}_{13} \slashed{\ep}(\wp)}{s_{13}}-\frac{\slashed{\ep}(\wp) \slashed{p}_{23} \slashed{\ep}(p_3)}{s_{23}} \right) 
  \, v(p_2) \,
\\ &=& \frac{\g}{g_e e_q} \SUNT^{\alpha_3}_{a_1 a_2} \, \Spamplitude^{(0)(a_1,a_2)}_{\gamma \to q_1 \bar{q}_2 \gamma_3}	\, ,
\label{SplittingLOA-qqbg}
\eeqn
while the polarized LO splitting function can be written as
\beqn
P^{(0),\mu \nu}_{\gamma \to q_1 \bar{q}_2 g_3} &=& e_q^2 g_e^2 \g^2 C_A C_F \,  {\cal P}^{\mu\nu}\left(p_1,p_2,p_3;\wp\right) \, .
\label{POLKERNELLOA-qqbgbis}
\eeqn

Centering in the NLO corrections, the divergent structure is dictated by
\beqn
\nn I^{(1)}_{\gamma \to q_1 \bar{q}_2 g_3} (p_1, p_2, p_3; \wp) &=& 
I^{(1)}_{\gamma \to q_1 \bar{q}_2 \gamma_3} (p_1, p_2, p_3; \wp) \nn \\ 
&+& \frac{\CG \g^2 C_A}{\ep^2} \left(\frac{-s_{123}-\imath 0}{\mu^2}\right)^{-\ep } \left(x_3^{-\ep}-x_1^{-\ep}-x_2^{-\ep} \right)~,
\eeqn
and
\beqn
c^{\gamma \to q \bar{q} g}&=& C_A C_F \, e_q^2 g_e^2 \g^4 \, ,
\label{NormalizacionAtoqqbarg}
\eeqn
is the global NLO normalization factor. As an usual check, we verified that all the $\ep$-poles were equal to those predicted by the expansion of $I^{(1)}_{\gamma \to q_1 \bar{q}_2 g_3}$.

Due to the presence of a non-trivial color structure, it is useful to decompose the ${\cal C}_{j}^{(i)}$ coefficients according to
\beqn
{\cal C}_{j}^{(i)} &=& C_A {\cal C}_{j}^{(i,C_A)} + D_A {\cal C}_{j}^{(i,D_A)} \, ,
\eeqn
where $D_A=C_F-C_A/2$ is related with the Abelian contributions. Moreover, we find that
\beqn
{\cal C}_{j}^{(i,D_A)} &=&{\cal C}_{j}^{(i,\gamma \to q \bar{q} \gamma)} \, ,
\eeqn
which was expected since the Abelian terms in $\gamma \to q \bar{q} g$ are the same that those present in the $\gamma \to q \bar{q} \gamma$ process. So, in order to simplify the presentation of the results, we only write the contributions proportional to $C_A$. The rational terms are given by
\beqn
{\cal C}_{1}^{(0,C_A)} &=& \frac{1-x_1}{2 x_1} \left(
\frac{8 (1-x_1)}{x_2} - 1 \right)~, \\ 
{\cal C}_{2}^{(0,C_A)} &=& - \frac{1}{1-x_1}~, \\ 
{\cal C}_{3}^{(0,C_A)} &=& \frac{1}{2 x_1\, x_2} \left( 
\frac{(1-x_2)^2}{1-x_1} + 15 - x_2  \right)~, \\
{\cal C}_{5}^{(0,C_A)} &=& \frac{1}{1-x_2} 
\left( \frac{1}{x_2} -\frac{1}{1-x_1} \right)~,
\eeqn
and
\beqn
{\cal C}_{1}^{(1,C_A)} &=& -\frac{3 (1-x_2)}{2 (1-x_1)} \log (x_1)~, \\
{\cal C}_{2}^{(1,C_A)} &=& - \frac{1}{(1-x_1)^2} 
\left( \frac{x_3}{x_1\, x_2} + 2 \right) \log (x_1)~, \\ 
{\cal C}_{3}^{(1,C_A)} &=& \frac{1}{2 x_1\,  x_2} \left( 
\frac{(1-x_2)^2}{(1-x_1)^2} \log (x_1)
- 3 \log (x_2) \right)~,
\\ {\cal C}_{5}^{(1,C_A)} &=& \frac{1}{1-x_1} \left( \frac{2(1-x_2)}{x_1\, x_2}
-\frac{1}{1-x_1} \right) \log (x_1)~, 
\eeqn
are the weight $1$ contributions. The non-trivial weight $2$ terms are given by
\beqn
{\cal C}_{1}^{(2,C_A)} &=& - \frac{(1-x_1)^2}{x_1 \, x_2} \, 
{\cal R}\left(x_1,x_2\right)~, \\ 
{\cal C}_{3}^{(2,C_A)} &=& - \frac{2}{x_1 \, x_2} \, {\cal R}\left(x_1,x_2\right)~.
\eeqn
It is interesting to appreciate that this is the last remaining $\cal R$-function involved in the expansion of standard scalar boxes. Also, we obtain the relation
\beqn
\sum_{j\in\{1,3\}} {\cal C}_{j}^{(2,C_A)} f_{j}^{\mu \nu} \, + \left(1 \leftrightarrow 2\right) &=& {\cal R}\left(x_1,x_2\right) \, \left.{\cal P}^{\mu \nu}\right|_{\epsilon^0} \, ,
\eeqn
which tells us that the weight $2$ contribution associated with $C_A$ is proportional to the LO splitting function.

Due to the fact that $\gamma \to q \bar{q} \gamma$ and  $\gamma \to q \bar{q} g$ share some Feynman diagrams in their perturbative expansion, the corresponding NLO corrections are related. This constitutes a cross-check of the results, since they were obtained from independent implementations. Explicitly, we have the relation
\beqn
P^{\mu \nu}_{\gamma \to q_1 \bar{q}_2 \gamma_3} &=& c^{\gamma \to q \bar{q} \gamma} \left(\left. \frac{P^{\mu \nu}_{\gamma \to q_1 \bar{q}_2 g_3}}{c^{\gamma \to q \bar{q} g}} \right|_{C_A\to 0}\right) \, ,
\eeqn
which is equivalent to cancel all the non-Abelian diagrams from $\gamma \to q \bar{q} g$ (and adapt the normalization due to the presence of an additional color matrix). The color structure of the LO splitting function $P^{\mu \nu}_{\gamma \to q_1 \bar{q}_2 g_3}$ is proportional to ${\rm Tr}\left[\SUNT^a \SUNT^a\right]=C_A C_F$ whilst $P^{\mu \nu}_{\gamma \to q_1 \bar{q}_2 \gamma_3} \propto {\rm Tr}\left[\IDC\right]$, using ${\rm Tr}\left[\IDC\right]=C_A$. In both processes, NLO corrections involve diagrams with virtual gluons, but only $\gamma \to q \bar{q} g$ allows triple-gluon vertices. Since these vertices are multiplied by a structure constant $f_{abc}$, they turn out to be proportional to $C_A$ when performing the contraction with the LO splitting amplitudes. As discussed in Ref. \cite{Sborlini:2014mpa}, virtual-gluon corrections with fermion-gluon vertices are multiplied by an additional factor $C_F$ or $D_A$ compared to the LO. In consequence, after factorizing the LO color structure, we can consider the limit $C_A \to 0$ to cancel diagrams with non-Abelian vertices. Moreover, in that limit, $C_F=D_A$ which leads to $P^{\mu \nu}_{\gamma \to q_1 \bar{q}_2 \gamma_3}$ after adding all the non-vanishing diagrams.\footnote{It is important to remark that this procedure is not the same as replacing the gauge group of the theory by an Abelian one. In an effective sense, it only accounts for the replacement of a gluon by a photon in a particular process.}

%20140730_1752: OK!!!

\subsection{Remarks on the structure of the photon-started splitting functions}
\label{sec:remarksphoton}
In order to make a proper analysis, let's recall the associated unpolarized results shown in Ref. \cite{Sborlini:2014mpa}. Before that, it is useful to introduce the notation
\beqn
\Delta_{i} &\equiv& x_i+z_i-1 \, ,
\eeqn
where the indices correspond to outgoing particles. For $\gamma \to q \bar{q} \gamma$ we found
\beqn
\nn \la \hat{P}_{\gamma \to q_1 \bar{q}_2 \gamma_3}^{(1)\,{\rm fin.}} \ra &=& \frac{C_F C_A}{2} e_q^4 g_e^4 \g^2 \, \left[ C^{(0)} + C_1^{(1)} \log(x_1)+  C_2^{(1)} \log(x_3) \right.
\\ &+& \left. \vphantom{C_1^{(1)}} C^{(2)} {\cal R}\left(x_1,x_3\right) +  \, (1 \leftrightarrow 2)\,\right]\, ,
\eeqn
with
\beqn
\nn C^{(0)} &=&  (x_1 x_2-z_1 z_2-{\Delta_1}{\Delta_2})\left(\frac{2-2 x_1 (x_2+1)}{x_1 x_2 (1-x_3)}-\frac{1}{1-x_1}\right) - \frac{8 (1-x_1)^2}{x_1 x_2}-\frac{1-x_1}{x_1}  
\\ &-& \frac{2 z_1 {\Delta_1} }{x_1 x_2}\left(\frac{(x_2+1) (1-x_2)^2}{(1-x_1) (1-x_3)}+\frac{x_3-x_1 x_2}{1-x_3}+\frac{(3-x_2) x_2-x_1 (x_2+1)}{2 (1-x_1)}+8\right) \, ,
\\ \nn C^{(1)}_{1} &=& \frac{x_1 x_2-z_1 z_2-{\Delta_1} {\Delta_2}}{(1-x_1) x_1}\left(\frac{x_2+2 x_3}{x_2^2}-\frac{x_1 x_2+2 x_3}{(1-x_1) x_2}-\frac{1+x_1}{1-x_1}\right)-\frac{z_2 (2 x_3-x_2) {\Delta_2}}{x_1 x_2^2}
\\ &-& \frac{(1-x_2)^2 z_1 (2 (1-x_1)+x_2) {\Delta_1}}{(1-x_1)^2 x_1 x_2^2}+\frac{(1-x_2) (x_2-2 x_3)}{(1-x_1) x_2} \, ,
\\ \nn C^{(1)}_{2} &=& \frac{2 \left(2 x_1 (z_2-1-{\Delta_2} (x_1 x_2+1))-{(\Delta^{0,3}_{1,2})}^2-2 x_2 (z_1+2 z_2-3) (x_1 x_2+2 z_3)-x_2 z_3\right)}{x_2^2 (1-x_3)^2}
\\ \nn &-&\frac{2 \left(x_1^2 (2 z_1 (z_2-3)+(4 z_2-13) z_2+7)+2 z_3^2\right)}{x_1 x_2 (1-x_3)^2} - 2 \left(\frac{z_3+x_1(z_2-1)}{x_2 (1-x_3)}\right)^2
\\ &-&\frac{2 (2 x_1 x_2+(z_1-15) z_1+7)}{(1-x_3)^2}  \, ,
\\ \nn C^{(2)} &=& \frac{2 x_2 \left(x_3 (z_1+x_3(z_2-1))+{\Delta_3} (2 z_1+x_2(z_3-1))+x_2^3+2 x_2 x_3 z_1\right)}{x_1 x_2^3} 
\\ &-& \frac{4 {\Delta_1} (x_3-z_1)}{x_1 x_2} + \frac{2 \left(z_1+x_3(z_2-1)\right)^2}{x_1 x_2^3} \, ,
\eeqn
for the finite NLO corrections, after applying a subtraction procedure analogous to the one described in \Eq{PpolarizedNLOdefinitionRENORMALIZED}. On the other hand, the corrections to $\gamma \to q \bar{q} g$ are given by
\beqn
\nn \la \hat{P}_{\gamma \to q_1 \bar{q}_2 g_3}^{(1)\,{\rm fin.}} \ra &=& \frac{D_A \g^2}{e_q^2 g_e^2} \la \hat{P}_{\gamma \to q_1 \bar{q}_2 \gamma_3}^{(1)\,{\rm fin.}} \ra + \frac{C^2_A C_F}{2} e_q^2 g_e^2 \g^4 \, \left[ C^{(0)} + C_1^{(1)} \log(x_1) \right.
\\ &+& \left. \KerLOA {\cal R}\left(x_1,x_2\right) +  \, (1 \leftrightarrow 2)\,\right]\, ,
\eeqn
with
\beqn
\nn C^{(0)} &=& \frac{16-7 x_2-2 z_1 z_2+(1-z_1)^2-15 z_2}{x_1}-\frac{z_1^2}{(1-x_1) x_2}-8\frac{z_1^2+(1-z_1)^2}{x_1 x_2}
\\ &+& \frac{2 z_1 (1-z_3)-x_2 (1-z_1)^2-(x_2+1) z_1}{(1-x_1) x_1}\, ,
\\ \nn C^{(1)} &=& \frac{{z_2} ({x_2} (4 {x_1} {z_1}+{x_1}-1)+2 {x_3} {z_1})+{x_2} ({x_1} (({x_2}-1) {z_1}+{x_2}-3)-2 {x_2}+3)}{({x_1}-1)^2 {x_1} {x_2}}
\\ &+& \frac{3 {x_2}^2+5 {x_2} ({z_2}-1)+3 {z_2}^2-4 {z_2}+1}{{x_1} {x_2}} -\frac{(1-{x_2})^2 {z_1}^2}{(1-{x_1})^2 {x_1} {x_2}}  \, ,
\eeqn
and
\beqn
\KerLOA &=& \frac{(\Delta_1)^2 + z_1^2}{2 x_1 x_2} \, + (1\leftrightarrow 2) = \left. \frac{\la \hat{P}_{q_1 \bar{q}_2 \gamma_3}^{(0)} \ra}{e_q^2 g_e^2 \g^2} \right|_{\ep=0}\, ,
\eeqn
which corresponds to the ${\cal O}\left(\ep^0\right)$ contribution to the $g \to q \bar{q} \gamma$ LO unpolarized splitting function.

It is interesting to appreciate that the coefficients ${\cal C}^{(i)}_{j}$ involved in the expansion of the polarized splitting functions are independent of the longitudinal-momentum fractions $z_i$, both for $P^{(1)\,{\rm fin.},\mu \nu}_{\gamma \to q_1 \bar{q}_2 \gamma_3}$ and $P^{(1)\,{\rm fin.},\mu \nu}_{\gamma \to q_1 \bar{q}_2 g_3}$. However, the unpolarized version of these splitting functions depends on $z_i$ in a non-trivial way. So, we conclude that these contributions are originated in the contraction of the different tensorial structures with the parent-gluon polarization tensor, $d_{\mu \nu}(\wp,n)$. Explicitly, we have
\beqn
d_{\mu \nu}(\wp,n) f_1^{\mu \nu} &=& -2(1-\ep) \, ,
\\ d_{\mu \nu}(\wp,n) f_2^{\mu \nu} &=& x_1 x_2- z_1 z_2- \Delta_1 \Delta_2 \, ,
\\ d_{\mu \nu}(\wp,n) f_3^{\mu \nu} &=& -2 z_1 \Delta_1 \, ,
\\ d_{\mu \nu}(\wp,n) f_{12}^{\mu \nu} &=& 0 \, ,
\eeqn
which also justifies the presence of the $\Delta_i$ functions in the final expressions, and $f_{12}^{\mu \nu}$ does not contribute because it is antisymmetric under the exchange $1 \leftrightarrow 2$ (or, equivalently, $\mu \leftrightarrow \nu$). Due to gauge invariance, photon-started splitting functions at loop-level can be computed using the replacement $d_{\mu \nu} \to -\eta^{\DST}_{\mu \nu}$ inside gluon propagators\footnote{For further details and a formal proof of this claim, see Ref. \cite{Sborlini:2014mpa}, Section IV.}. If we remove the polarization vector associated with the parent parton, then it is possible to compute the splitting amplitude without explicitly taking into account the LCG quantization vector $n^{\mu}$. This property is straightforwardly translated into $P^{\mu \nu}$, because this object is computed using the product of amputated splitting amplitudes. Thus, the coefficients ${\cal C}^{(i)}_j$ for the collinear processes $\gamma \to a_1 \ldots a_m$ must be independent of $z_i$ (and, of course, $n \cdot \wp$).

Anyway, as we discussed in Ref. \cite{Sborlini:2014mpa}, photon-started splitting processes can be computed without performing any $\epsilon$-expansion because they only involve standard boxes and bubbles that are known in terms of exponentials and hypergeometric functions. Since the exact expressions are lengthy, we present them in Appendix A.

%***********************************************************************************************************************************************%
%***********************************************************************************************************************************************%
\section{Gluon-started splitting: $g \to q \bar{q} \gamma$}
\label{sec:gluon}
%***********************************************************************************************************************************************%
%***********************************************************************************************************************************************%
Finally, we arrive to the gluon started splitting. Due to the fact that it is originated from a parent gluon, there is a non-trivial color flow and it is not possible to remove all LCG integrals to trivially avoid all $z_i$ dependence, as happened in the $\gamma \to a_1 \ldots a_m$ processes.

Starting with the tree-level contributions, the splitting amplitude is
\beqn
\Spamplitude^{(0)(a_1,a_2;\alpha)}_{q_1 \bar{q}_2 \gamma_3} &=& \frac{e_q g_e \g \mu^{2\epsilon} \SUNT^{\alpha}_{a_1 a_2}}{ {s_{123}}} \, \bar{u}(p_1) 
\left( \frac{\slashed{\epsilon}(p_3) \slashed{p}_{13} \slashed{\epsilon}(\wp)}{s_{13}}-\frac{\slashed{\epsilon}(\wp) \slashed{p}_{23} \slashed{\epsilon}(p_3)}{s_{23}} \right) 
  \, v(p_2) \, \,
\, ,
\label{SplittingLOg-qqbA}
\eeqn
and the polarized splitting function is given by
\beqn
P^{(0),\mu \nu}_{q_1 \bar{q}_2 \gamma_3} &=& \frac{e_q^2 g_e^2 \g^2}{2} \,  {\cal P}^{\mu\nu}\left(p_1,p_2,p_3;\wp\right) \, .
\label{POLKERNELLOg-qqbAbis}
\eeqn
Analysing the NLO corrections to this process, the normalization factor is given by
\beqn
c^{q \bar{q} \gamma} &=& \frac{e_q^2 g_e^2 \g^4}{2} \, ,
\eeqn
and the divergent structure is in complete agreement with the one predicted by Catani's formula, i.e.
\beqn
\nn I^{(1)}_{q_1 \bar{q}_2 \gamma_3} (p_1, p_2, p_3; \wp) &=& \frac{\CG \g^2}{\ep^2} \left(\frac{-s_{123}-\imath 0}{\mu^2}\right)^{-\ep } \left[\vphantom{\frac{\beta_0}{2}}C_A \left(2-z_1^{-\ep}-z_2^{-\ep}+x_3^{-\ep}\right)
\right. \\ &-& \left. 2 C_F \, x_3^{-\ep} - \ep \left(2\gamma_q -\gamma_g - \frac{\beta_0}{2}\right)\right] \, .
\label{ICgqqbANLO}
\eeqn
Notice that the $\ep$-expansion of \Eq{ICgqqbANLO} involves the presence of $\log^j (z_i)$  ($i,j=1,2$), which implies that it could be possible to have some $z_i$ dependence in $P^{(1)\,{\rm fin.},\mu \nu}_{q_1 \bar{q}_2 \gamma_3}$.

As we did previously, it is convenient to classify the different color contributions to ${\cal C}^{(i)}_j$. So, we use the decomposition
\beqn
{\cal C}_{j}^{(i)} &=& C_A {\cal C}_{j}^{(i,C_A)} + D_A {\cal C}_{j}^{(i,D_A)}+  \beta_0 {\cal C}_{j}^{(i,\beta_0)} \, ,
\eeqn
with
\beqn
{\cal C}_{j}^{(i,D_A)} &=&{\cal C}_{j}^{(i,\gamma \to q \bar{q} \gamma)} \, ,
\eeqn
because the Abelian component of this splitting function coincides with $P^{(1), \mu\nu}_{\gamma \to q \bar{q} \gamma}$. Applying this notation, the terms proportional to $\beta_0$ are expressed as 
\beqn
\left. P^{(1)\,{\rm fin.},\mu \nu}_{q_1 \bar{q}_2 \gamma_3} \right|_{\beta_0} &=& \frac{10}{3} \, c^{q \bar{q}\gamma} \, {\cal P}^{\mu \nu} \, ,
\eeqn
because this contribution is originated from the self-energy correction of the parent gluon \cite{Sborlini:2013jba}. 

After these appreciations, we need to present only ${\cal C}_{j}^{(i,C_A)}$ to complete the description of the $g \to q \bar{q} \gamma$ splitting function. The rational terms are given by
\beqn
{\cal C}_{1}^{(0,C_A)} &=& \frac{1-x_1}{2 x_1} \left( 
\frac{10\, (1-x_1)}{3 x_2} + 1 \right)~, \\ 
{\cal C}_{2}^{(0,C_A)} &=& \frac{1}{1-x_1}~, \\ 
{\cal C}_{3}^{(0,C_A)} &=& - \frac{1}{2 x_1\, x_2}
\left( \frac{(1-x_2)^2}{1-x_1} - \frac{23}{3} - x_2  \right)~, 
\\ {\cal C}_{5}^{(0,C_A)} &=& \frac{1}{x_1 (1-x_1)} \, ,
\eeqn
while the contributions of weight $1$ are
\beqn
{\cal C}_{1}^{(1,C_A)} &=& - \frac{3 (1-x_2)}{2 (1-x_1)} \log (x_1)~, \\ 
{\cal C}_{2}^{(1,C_A)} &=& - \frac{1}{(1-x_1)^2} \left(
\frac{2 x_3}{x_1\, x_2} + 1 \right) \log (x_1)~, \\ 
{\cal C}_{3}^{(1,C_A)} &=& - \frac{1}{2 \, x_1\, x_2} \left(
\frac{(1-x_2)^2}{(1-x_1)^2}  \log (x_1) + 3 \log (x_2) \right) ~, 
\\ {\cal C}_{5}^{(1,C_A)} &=& \frac{1}{(1-x_1)^2} \left(
\frac{x_3}{x_1\, x_2} + 2 \right) \log (x_1)~.
\eeqn
As we could appreciate for the photon-started splitting processes, all the contributions were independent of $z_i$ due to the lack of LCG integrals. However, the same behaviour is observed here, at least for weights $0$ and $1$. In this case, a cancellation among the $z_i$-logarithms in $P^{(1),\mu\nu}_{q_1 \bar{q}_2 \gamma_3}$ and those in $I^{(1)}_{q_1 \bar{q}_2 \gamma_3}$ takes place.

The situation changes when studying weight $2$ contributions, which are more complicated than in the previous splitting functions. For this reason, a more sophisticated procedure was required for their treatment. The first step consisted in identifying a set of functions to expand these terms. Following the choice shown in Ref. \cite{Sborlini:2014mpa} for the unpolarized splitting function $g \to q \bar{q} \gamma$, we have the basis
\beqn
\nn F_1 &=& \frac{\pi ^2}{6}-2 \li{1-\frac{x_1}{1-z_1}}-2\li{1-\frac{z_2}{1-z_1}}+2 \li{1-z_1}
\\ &+& 2 \log (x_2) \log (1-z_1)\, + (1\leftrightarrow 2) ,
\\ F_2 &=& \log (x_1) \log (x_2) \, ,
\\ F_3 &=& \frac{\pi ^2}{4}-\li{1-x_1}-\log (x_1) \log (z_1) + (1\leftrightarrow 2) \, ,
\\ F_4 &=& \log \left(\frac{x_1}{1-z_1}\right) \log \left(\frac{1-z_1}{z_1 z_2}\right)-\log (x_2) \log (1-z_1) \, ,
\\ F_5 &=&  \log \left(\frac{x_2}{1-z_2}\right) \log \left(\frac{1-z_2}{z_1 z_2}\right)-\log (x_1) \log (1-z_2) = S_{1\leftrightarrow 2} \left(F_4\right) \, ,
\eeqn
whose associated coefficients are
\beqn
\nn {\cal C}_{1}^{(2,C_A)} &=& -\frac{1}{2 x_1\, x_2} \Bigg[ \frac{F_1}{2} 
+ \left((x_2-x_1) \frac{x_3+(1-x_1) \Delta_2}{\Delta_1}
-\frac{(1-x_3)^2}{2} \right) F_2  - 2 (1-x_1)^2 F_3 
\\ &+& \left((x_1-x_2) \left(\frac{z_3+x_1 \Delta_2}{\Delta_1}-x_2\right) -\frac{x_1 \, x_3+(1-x_3) \Delta_1}{z_3}-x_1-4 x_2+3\right) F_4 \Bigg]~,
\\ \nn {\cal C}_{2}^{(2,C_A)} &=& \frac{(x_1\, z_1+x_2 \, z_2+z_3) (2 z_1 z_2-x_1 (1-z_2)-x_2 (1-z_1)+z_3)}
{2 x_1\, x_2} \frac{F_1}{\Omega} - \bigg(\frac{(x_3 + (1-x_1) \Delta_2)^2}{x_2 \, \Delta_1} 
\\ \nn &+& \frac{x_3 \left( 1 + 2 z_2 - 2 (z_3 + 2 z_2) z_1\right)+z_2 (2 z_1-2 z_2-1)}{x_1} -x_1 (2+\Delta_2) + z_3 + 4 z_1^2 \bigg) \frac{F_2}{\Omega} 
\\ \nn &+& \bigg[\frac{(1-z_1)^2 \Delta_1-x_1 \, z_3}{x_2\, z_1\, z_3\, \Delta_1} + \frac{z_2}{\Omega} \bigg(\frac{x_1 (1-z_2) \left(2 z_1^2-(1-z_2) z_2\right)}{x_2 \, z_3} - \frac{x_2 \left(1-z_1^2\right) z_2}{x_1 \, z_3}
\\ \nn &-& \frac{2 \left((1-z_1) z_1^2+(1-2z_1-z_2) z_2\right)}{z_3} + \frac{(2 z_1-z_2) z_3}{x_1 \, x_2} + \frac{2 (z_2-(1-z_1) z_1)}{x_1} 
\\ &+& \frac{2 (z_1 + z_1^2 +z_2 z_3)}{x_2} \bigg) \bigg] F_4~,
\\ \nn {\cal C}_{3}^{(2,C_A)} &=& - \frac{1}{2 x_1\, x_2} \left(\frac{2 (x_1\, z_1+x_2\, z_2+z_3) z_2 \, \Delta_2}{\Omega} 
+ 1\right) F_1 + \bigg(\frac{z_2 (z_2-x_3 (2z_2+z_3))}{x_1}   
\\ \nn &+& \frac{(x_1 (x_2+z_2)-z_2)\left(x_1^2\, x_2+x_2\, z_2^2-x_1\, z_2\, \Delta_2\right)}{x_1\, x_2\, \Delta_1} -\frac{(1-z_2) (x_3 (1-z_2)-z_1+z_2)}{x_2}
\\ \nn &+& 1 - x_2 (x_1+z_3) - (x_1+2) z_2 -2 (1-z_2) z_1 \bigg) \frac{F_2}{\Omega} + \frac{2}{x_1\, x_2}  F_3
\\ \nn &+& \bigg(\frac{(x_1 (x_2+z_2)-z_2) \left(x_1^2 x_2+x_2 z_2^2-x_1\, z_2\, \Delta_2 \right)}{x_1\, x_2 \Delta_1} -\frac{(2-z_2) z_3^2}{x_1 x_2} - \frac{2 (1-x_3) z_1 z_2^3}{x_1 z_3}
\\ \nn &+& \frac{z_3^2 (x_3 z_1-z_2)+(1-z_1) (z_3 (4-3 x_3 z_1+x_3)+(1-z_1) (2 x_3 z_1+x_3-2))}{x_1} + 1
\\ \nn &+& \frac{(1-z_2) \left(x_3 (1-z_2) + 2 - 3 z_1 - 2 z_2 + 3 z_1 z_2-z_2^2\right)}{x_2} +\frac{2 (1-z_2) z_2^2}{z_3} - x_2 (x_1+z_3)
\eeqn
\beqn
\nn &-& (x_1+z_1^2+z_2^2) z_2 - 2 (1-z_3) z_3 \bigg) \frac{F_4}{\Omega}-\bigg[ \frac{z_2}{\Omega \, x_1}\left(\frac{(2 z_1+z_3) (x_1 z_1-x_3 z_3)}{x_2} + z_1 z_3 \right.
\\ &+& \left. \frac{1}{z_3}\left(x_1 \left(z_1^2 (z_2+1)-z_1 (1-z_2)+(1-z_2)^2 z_2\right)- \, (1\leftrightarrow 2) \right)\right)+\frac{1}{x_1 \, x_2} \bigg] F_5~,
\\ {\cal C}_{5}^{(2,C_A)} &=& -\frac{F_2+F_4}{x_1\, \Delta_1} - \frac{(1-z_1)^2+(1-z_2)^2}{x_1\, x_2\, z_3} F_4~.
\eeqn
For these contributions, there is a non-trivial dependence in both $z_i$ and $\Delta_i$, not only inside the rational coefficients but also in the definition of the transcendental functions $F_i$. This is a consequence of the presence of Feynman integrals with LCG denominators, which is closely related to non-Abelian interactions.

As a final comment, let's discuss about the possible functional dependence of the rational coefficients. The description of the triple collinear limit involves three \textit{almost-collinear} momenta, $p_i^{\mu}$, and the quantization direction $n^{\mu}$. Since we are computing scalar objects, they can only depend on the scalar products $n \cdot p_i$ and $s_{i j}$. Moreover, the result is dimensionless and independent of $n^{\mu}$, which justifies the introduction of the variables $z_i$ and $x_i$. But these variables are not independent, i.e. they fulfil
\beq
\sum_{i=1}^3 x_i = 1 \ \ \ \ , \ \ \ \ \sum_{i=1}^3 z_i = 1\, ,
\eeq
which implies that we can describe all the results using four variables. Motivated by the possibility of simplifying the expressions, we introduced the variables $\Delta_i$ that correspond to the scalar products of the collinear direction $\wp^{\mu}$ and the momentum $p_{i}^{\mu}$; explicitly,
\beqn
\nn 2 \wp \cdot p_i &=& s_{123} \left(\frac{2 p_i \cdot p_j + 2 p_i \cdot p_k}{s_{123}} - \frac{n \cdot p_i}{n \cdot \wp} \right) = s_{123} \left(\frac{s_{123} - s_{jk}}{s_{123}} - \frac{n \cdot p_i}{n \cdot \wp} \right)
\\ &=& s_{123} \left(1-x_i-z_i\right) = - s_{123} \, \Delta_i \, ,
\eeqn
where we considered $i \neq j \neq k$. For this reason, the variables $\Delta_i$ have a well-defined physical meaning and it is expected that they appear in the calculation. Since weight 0 and 1 contributions depend only on $x_i$, we could replace $z_i = 1- x_i + \Delta_i$ and rewrite all the coefficients ${\cal C}_j$. However, we decided to mix the different variables involved in the problem with the purpose of reducing the length of the final results.

\subsection{Comments on cross-checks}
As we did with all the previous processes, the first check consisted in comparing the divergent structure with Catani's formula. In this particular case, we carefully studied the cancellation of higher weight functions that were multiplying single $\ep$-poles. Since we are performing operations with matrices whose elements have $\ep$-poles ($M^{-1}$, as defined in Section \ref{sec:details}), some transcendental weight $2$ functions associated with the finite pieces of Feynman integrals could contribute to the divergent IR structure. Of course, Catani's formula rules out this possibility. However, we explore this issue putting flags in some integrals. Explicitly, the triple collinear limit involves the massless box-integral
\beqn
I^{\rm box}_{\rm LCG} &=& \int_q \, \frac{1}{\left(q^2+\imath 0\right)\left((q-p_2)^2+\imath 0\right)\left((q-p_{23})^2+\imath 0\right)\left((q-p_{123})^2+\imath 0\right) \, (n \cdot q+\imath 0)} \, ,
\eeqn
which is known up to order $\epsilon^0$. If we perform a naive general $\ep$-expansion, we have
\beqn
I^{\rm box}_{\rm LCG} &=& \CG \, \g^2 \, \left( \frac{-s_{123} -i0}{\mu^2}\right)^{-\ep} \, \left(\frac{B_0}{\epsilon^2} + \frac{B_1}{\epsilon} + B_2 \right) \, ,
\eeqn
where $B_0$ only contains rational functions and $B_i$ incorporates transcendental functions of weight up to $i$. So, we studied the cancellation of single $\ep$-poles without writing down the explicit form of $B_2$. Since subtracting $I^{(1)}_{q_1 \bar{q}_2 \gamma_3}$ removes all the divergences, we obtained the following equation
\beqn
\frac{1}{\ep} \left[\frac{B_2 + S_{1 \leftrightarrow 2}\left(B_2\right)}{2} + {\cal D}(x_i,z_i) \right] &=& 0 \, ,
\eeqn
with ${\cal D}(x_i,z_i)$ only involves rational combinations of weight $2$ functions. In consequence, this procedure allowed us to perform a cross-check among our polarized splitting results and the ${\cal O}\left(\epsilon^0\right)$ terms of LCG Feynman integrals, which were computed using other methods.

Following a more conventional path, we also checked that the final result is symmetric under the exchange $1 \leftrightarrow 2$. Another test consisted in taking the limit $C_A \to 0, N_f \to 0$ of the normalized splitting function and comparing it with $P^{\mu\nu}_{\gamma \to q \bar{q} \gamma}$. Explicitly, the relation
\beqn
P^{\mu \nu}_{\gamma \to q_1 \bar{q}_2 \gamma_3} &=& c^{\gamma \to q \bar{q} \gamma} \left(\left. \frac{P^{\mu \nu}_{q_1 \bar{q}_2 \gamma_3}}{c^{q \bar{q} \gamma}} \right|_{C_A\to 0, N_f\to 0}\right) \, ,
\eeqn
turns out to be successfully verified. Notice that, in addition to the limit $C_A \to 0$ described in Section \ref{sec:photon}, here we also require $N_f \to 0$. This leads to the complete cancellation of gluon self-energy corrections and it allows a one-to-one correspondence between the Feynman diagram expansion of $\gamma \to q \bar{q} \gamma$ and $g \to q \bar{q} \gamma$.

Finally, we contracted $P^{(1)\,{\rm fin.},\mu\nu}_{\gamma \to q \bar{q} \gamma}$ with $d_{\mu \nu}(\wp,n)$ to recover the unpolarized splitting function (which was computed with an independent implementation). Again, we found a complete agreement.

%***********************************************************************************************************************************************%
\section{Conclusions}
\label{sec:conclusion}
%***********************************************************************************************************************************************%
In this paper, we computed all the relevant polarized splitting functions in the triple collinear limit, for processes involving at least one photon: $\gamma \to q \bar{q} \gamma$, $\gamma \to q \bar{q} g$ and $g \to q \bar{q} \gamma$. We obtained the NLO corrections to these objects, working in CDR and using TL-kinematics, where strict collinear factorization is fulfilled.

Due to gauge invariance, photon-started polarized splitting functions are completely independent of $n^{\mu}$ and independent of the longitudinal-momentum fractions $z_i$ too. This reduces the amount of transcendental functions required to express the results. Moreover, weight $2$ components are very simple because they turn out to be proportional to the function ${\cal R}(x_i,x_j)$. 

The fact that $P^{\mu \nu}_{q_1 \bar{q}_2 \gamma_3}$ is a gluon-initiated process implies a rather different behaviour of this splitting function compared with the others. In particular, LCG Feynman integrals are required for the computation. Also, the components of transcendental weight $2$ depend on $z_i$ and $\Delta_i$. However, all this contributions are isolated in terms proportional to $C_A$, because the Abelian part is related to $P^{\mu \nu}_{\gamma \to q_1 \bar{q}_2 \gamma_3}$.

All the results shown in this article were compared against their unpolarized version, presented in Ref. \cite{Sborlini:2014mpa}, and they were consistent. Besides that, we implemented some cross-checks among the polarized splitting functions, in particular, testing the limit $C_A \to 0, N_f \to 0$ after removing the LO normalization. An alternative test was proposed to check the $g \to q \bar{q} \gamma$ splitting function. Relying on Catani's formula, it is expected that single $\ep$-poles do not contain any weight $2$ function. On the other hand, the ${\cal O}(\epsilon^0)$ pieces of all the LCG integrals involved contain only this kind of functions. So, we took the LCG massless box and replaced the finite piece with a generic expression. Then, we forced the cancellation of single $\ep$-poles and obtained an additional constraint which relates Feynman integrals expansions and polarized splitting functions. Due to the fact that they were computed independently, this comparison provides another check to our results.

Finally, we would like to emphasize that NLO corrections to polarized splitting functions in the triple collinear limit are essential ingredients for NNNLO computations and beyond. Pure QCD triple-splitting processes, which have a more complicated color structure, will be presented in a forthcoming article.

%***********************************************************************************************************************************************%

%%%%%%%%%%%%%%%%%%%%%%%%%%%%%%%%%%%%%%%%%%%%%%%%%%%%%%%%%%%%%%%%%%%%%%%  
  \subsection*{Acknowledgements}
This work is partially supported by UBACYT, CONICET, ANPCyT, the
Research Executive Agency (REA) of the European Union under
the Grant Agreement number PITN-GA-2010-264564 (LHCPhenoNet),
by the Spanish Government and EU ERDF funds
 (grants FPA2011-23778 and CSD2007-00042
Consolider Ingenio CPAN) and by GV (PROMETEUII/2013/007).
%%%%%%%%%%%%%%%%%%%%%%%%%%%%%%%%%%%%%%%%%%%%%%%%%%%%%%%%%%%%%%%%%%%%%%%  

%%%%%%%%%%%%%%%%%%%%%%%%%%%%%%%%%%%%%%%%%%%%%%%%%%%%%%%%%%%%%%%%%%%%%%%
\section*{Appendix A: Exact results for photon-started processes}
Here we present the expressions for the photon-started splitting functions in terms of boxes and bubbles, without performing any $\epsilon$-expansion or $\epsilon$-pole subtraction. We write the functions according to
\beqn
P^{(1)\,,\mu \nu}_{\gamma \to a_1 a_2 a_3} &=& c^{\gamma \to a_1 a_2 a_3} \left[ \sum_{j=1}^{4} \, A^{(1)}_j f_{j}^{\mu \nu} \, + \, A^{(1)}_5 f_{12}^{\mu \nu} \right] \, ,
\label{EquacionDescomposicionEXACTA1}
\eeqn
where the coefficients $A^{(1)}_j$ are expressed as linear combinations of master integrals (MI) multiplied by rational functions that depend only on $x_i$ and $\epsilon$.

\subsection*{$\gamma \to q \bar{q} \gamma$}
The normalization factor for this process is defined in \Eq{NormalizacionAtoqqbarA} and the expansion of the coefficients introduced in \Eq{EquacionDescomposicionEXACTA1} involves only the MIs presented in Eqs. (\ref{DEFBoxes}-\ref{DEFBubbles}). After taking into account symmetry considerations, we obtain
\beqn
\nn A^{(1)}_1 &=& \frac{I^{({\rm box})}_{13} x_3 }{x_2} \left(\frac{2 \left(\epsilon ^3+3 \epsilon ^2-\epsilon +1\right) \left(1-x_3\right)}{(1-2 \epsilon) x_1^{-1} }+\frac{(\epsilon -1) \left(x_1+5 x_2-1\right)}{x_1^{-1} x_2 (1-2 \epsilon )}-x_2^2 \epsilon +(1-x_2)^2+1\right) 
\\ \nn &+& I^{({\rm bubble})}_0 \left(\frac{2 \left(\epsilon ^2-2 \epsilon +2\right)}{x_1 x_2 \epsilon }+\frac{x_2 \left(2 \epsilon ^3+15 \epsilon ^2-7 \epsilon +6\right)-2 \left(2 \epsilon ^2-5 \epsilon +6\right)}{x_1 \epsilon } -\frac{2}{1-x_3}\right.
\\ \nn &-& \left. \frac{(2 \epsilon +3) (x_1 (\epsilon-1)+\epsilon +1)}{1-x_2}+\frac{2 x_3 (1-\epsilon )}{x_2^2 \epsilon }+\frac{4 \left(\epsilon ^3+3 \epsilon ^2-\epsilon +1\right)}{\epsilon }\right)
\\ \nn &+& I^{({\rm bubble})}_1 \left(\frac{x_2 \left(\epsilon ^2-8 \epsilon +4\right)-x_2^2 \left(\epsilon ^3+4 \epsilon ^2-6 \epsilon +2\right)+8 \epsilon -4}{x_1 x_2 \epsilon } \right.
\\ \nn &-& \left. \frac{2 x_1 \left(2 x_2 \left(\epsilon ^3+3 \epsilon ^2-\epsilon +1\right)+\epsilon -1\right)}{x_2^2 \epsilon }+\frac{(2 \epsilon +3) (x_2 (\epsilon -1)+\epsilon +1)}{1-x_1} \right.
\\ \nn &-& \left. \frac{x_2^2 \left(\left(7 \epsilon ^3+16 \epsilon ^2-6 \epsilon +4\right)\right)+2 x_2 (6 \epsilon -5)-2 (\epsilon -1)}{x_2^2 \epsilon }\right)
\\ \nn &+& I^{({\rm bubble})}_3 \left(\frac{2 (x_2-1) (1-\epsilon ))}{x_1^2 \epsilon }-\frac{2 \left(2 x_2^2 \left(\epsilon ^3+3 \epsilon ^2-\epsilon +1\right)+x_2 (4 \epsilon -5)-3 \epsilon +2\right)}{x_1 x_2 \epsilon } \right.
\\ &-& \left. \frac{4 x_1 x_2 \left(\epsilon ^3+2 \epsilon ^2+1\right)+3 \epsilon -2}{x_1 x_2 \epsilon }+\frac{2}{1-x_3}\right) + \left(1 \leftrightarrow 2\right)\, ,
\\ \nn A^{(1)}_2 &=& \frac{I^{({\rm box})}_{13} }{1-2 \epsilon } \left(\frac{2 (x_1-1)^2 (\epsilon -1)^2}{x_2^3}-2 \left(\epsilon  \left(\epsilon  (\epsilon +5)+x_1 \left(\epsilon ^2+\epsilon +2\right)-2\right)+1\right) \right.
\\ \nn &-& \left. \frac{2 \left((x_1 (x_1+2)-1) \epsilon ^3+(x_1 (x_1+7)-6) \epsilon ^2+(x_1 (2 x_1-3)+4) \epsilon +x_1-3\right)}{x_2} \right.
\\ \nn &-& \left. \frac{2 (x_1-1) \left(\epsilon  \left(x_1 (\epsilon +1)^2-2 \epsilon +4\right)-3\right)}{x_2^2}\right) + I^{({\rm bubble})}_0 \left(\frac{2 (\epsilon +1)^2}{x_2^2}-\frac{2 (\epsilon -1)^2}{(1-x_1)^2 x_1^3 \epsilon } \right. 
\eeqn
\beqn
\nn &+& \left. \frac{2 (1-\epsilon )}{x_1^2 x_2^2}+\frac{1-\epsilon ^2}{1-x_2}+\frac{\epsilon ^2+2 \epsilon +5}{x_2} -\frac{3 \left(\epsilon ^2+\epsilon +1\right)}{(1-x_1) x_1^3 x_2}-\frac{2 (2 \epsilon  x_1-3 x_1+2)}{x_1^3 (1-x_3)} \right.
\\ \nn &+& \left. \frac{2 x_1 \epsilon ^2-\epsilon ^2+2 x_1 \epsilon +2 \epsilon +4 x_1-9}{(1-x_1)^2}+\frac{2 \left(\epsilon ^3+4 \epsilon ^2-3 \epsilon +2\right)}{(1-x_1)^2 x_1^2 \epsilon } - \frac{3 \left(\epsilon ^2+\epsilon +1\right)}{x_1 (1-x_2)}\right.
\\ \nn &+& \left. \frac{3 \epsilon ^3+17 \epsilon ^2-9 \epsilon +4}{x_1 x_2 \epsilon }+\frac{\epsilon  (\epsilon  (3 \epsilon +7)-1)-2 x_2 (\epsilon -1)^2+4}{x_1^3 x_2 \epsilon } - \frac{4 \left(x_1^2+1\right)}{x_1^2 (1-x_3)^2} \right. 
\\ \nn &-& \left. \frac{3 \epsilon ^3+8 \epsilon ^2-5 \epsilon +2}{(1-x_1)^2 x_1 \epsilon }-\frac{8-\epsilon  (3 \epsilon  (\epsilon +1)+5)}{x_1^2 x_2 \epsilon }\right) + 2 I^{({\rm bubble})}_1 \left(\frac{2 (x_1-1) (1-\epsilon )^2}{x_1 x_2^3 \epsilon } \right.
\\ \nn &+& \left. \frac{3 \left(\epsilon ^2+\epsilon +1\right)}{(1-x_1) x_2}-\frac{(\epsilon -1) \left(\epsilon  x_1^2+2 x_1-\epsilon -1\right)}{(x_1-1)^2 x_1} - \frac{3 \epsilon ^3+11 \epsilon ^2-7 \epsilon +2}{x_1 x_2 \epsilon } \right.
\\ \nn &-& \left. \frac{2 \left(\epsilon ^2+\epsilon +2\right)}{x_2} -\frac{2 \left(x_1 \epsilon ^3+2 x_1 \epsilon ^2-2 \epsilon ^2+x_1 \epsilon +3 \epsilon -2\right)}{x_1 x_2^2 \epsilon }\right) + I^{({\rm bubble})}_3 \left(\frac{4 (1-\epsilon )^2}{x_1^3 \epsilon } \right.
\\ \nn &+& \left. \frac{2 (\epsilon -1)}{x_1^2 x_2^2}+\frac{2 \epsilon  (\epsilon +1)}{x_1 x_2 x_3}+\frac{8 (\epsilon +1)}{x_1 x_2^2 \epsilon }-\frac{4 \left(\epsilon ^2+\epsilon +2\right)}{x_1}-\frac{4 (\epsilon +1)^2}{x_1^2} - \frac{6}{x_1 x_2^2}+\frac{4}{x_1 x_2^3} \right.
\\ \nn &-& \left. \frac{4 \epsilon }{1-x_3}+\frac{2 (2-3 x_1)}{x_1^3 (1-x_3)}+\frac{4}{(1-x_3)^2}+\frac{4}{x_1^2 (1-x_3)^2}-\frac{2 (\epsilon  (\epsilon  (3 \epsilon +7)-1)+2)}{x_1 x_2 \epsilon } \right.
\\ &-& \left. \frac{4 \left(\epsilon ^2+1\right)}{x_1^3 x_2 \epsilon }\right) \, + \left(1 \leftrightarrow 2\right) \, ,
\\ \nn A^{(1)}_3 &=&  I^{({\rm box})}_{13} \left(\frac{(1-x_1) (\epsilon -1)^2}{x_2^3 (1-2 \epsilon )}+\frac{x_2 \epsilon  \left(\epsilon ^2+\epsilon +2\right)}{1-2 \epsilon } -\frac{\epsilon  (3 \epsilon -4)-x_1 (\epsilon  (\epsilon  (\epsilon +3)-2)+2)+3}{x_2^2 (1-2 \epsilon )}\right.
\\ \nn &-& \left. \frac{3-\epsilon  \left(x_1 \left(\epsilon ^2+\epsilon +2\right)+2-\epsilon  (2 \epsilon +7)\right)}{1-2 \epsilon }+\frac{\epsilon  (\epsilon  (\epsilon +8)-6)-x_1 (3-\epsilon  (\epsilon +4) (1-2 \epsilon ))+5}{x_2 (1-2 \epsilon )}\right)
\\ \nn  &+& I^{({\rm box})}_{23} \left(\frac{(\epsilon -1)^2 (1-x_2)^3}{x_1^3 (1-2 \epsilon )}+\frac{\left(\epsilon  \left(-\epsilon +x_2 \left(\epsilon ^2+\epsilon +2\right)+4\right)-3\right) (1-x_2)^2}{x_1^2 (1-2 \epsilon )}+2 (\epsilon -1) \right.
\\ \nn &-& \left. \frac{(x_2-1) \left(\epsilon  \left(-\epsilon  (\epsilon +5)+x_2 \left(\epsilon ^2+\epsilon +2\right)+8\right)-4\right)}{x_1 (2 \epsilon -1)}\right) + I^{({\rm bubble})}_0 \left(\frac{2 x_2 (1-\epsilon )^2}{x_1^3 \epsilon }-\frac{4 (1-\epsilon )^2}{x_1^3 \epsilon } \right. 
\\ \nn &+& \left. \frac{2 (1-\epsilon )^2}{x_1^3 x_2 \epsilon }+\frac{x_2 (1-\epsilon )}{(1-x_1)x_1}+\frac{2 (1-\epsilon )}{x_1^2 x_2^2} - \frac{2 \epsilon }{1-x_1}+\frac{(1-\epsilon)x_2^2-2 x_2-\epsilon +1}{(1-x_1)^2 x_2} \right. 
\\ \nn &+& \left. \frac{2 \left(\epsilon ^2+1\right)}{x_1 x_2^3 \epsilon }+\frac{2 \left(\epsilon ^2+\epsilon -2\right)}{x_1^2 x_2 \epsilon } - \frac{2 \epsilon ^2-\epsilon +3}{(1-x_1) x_2}+\frac{2 \left(11 \epsilon ^2-8 \epsilon +5\right)}{x_1 x_2 \epsilon }+\frac{2 \left(\epsilon ^3+\epsilon ^2+2\right)}{x_1^2 \epsilon } \right. 
\\ \nn &-& \left. \frac{2 \left(\epsilon ^2+2 \epsilon +2\right)}{x_1}-\frac{2 x_2 \left(\epsilon ^2+\epsilon +2\right)}{x_1^2}+\frac{2 \left(\epsilon ^2+1\right)}{(1-x_1) x_2^2}-\frac{4 \left(x_2^2+\epsilon  x_2-2 x_2+1\right)}{x_2^3 (1-x_3)} \right. 
\\ \nn &-& \left. \frac{4 \left(x_2^2-x_2+1\right)}{x_2^2 (1-x_3)^2}-\frac{2 \left(\epsilon ^2+\epsilon +2\right)}{x_1 x_2^2 \epsilon }\right) + I^{({\rm bubble})}_1 \left(-\frac{x_2 (1-\epsilon)^2}{1-x_1}-\frac{2 (1-\epsilon )^2}{x_1 x_2^3 \epsilon }+\frac{\epsilon ^2+3}{(1-x_1) x_2} \right. 
\\ \nn &+& \left. \frac{\epsilon  x_2^2-x_2^2+2 x_2+\epsilon -1}{(1-x_1)^2 x_2}-\frac{2 \left(\epsilon ^2+1\right)}{(1-x_1) x_2^2} + \frac{2 \left(\epsilon ^2+\epsilon +2\right)}{x_1}+\frac{2 \left(3 \epsilon ^2-3 \epsilon +2\right)}{x_1 x_2^2 \epsilon } -\frac{x_2 (\epsilon -1)^2}{x_1} \right.
\\ \nn &-& \left. \frac{3 \epsilon ^3+14 \epsilon ^2-11 \epsilon +6}{x_1 x_2 \epsilon }\right) + \frac{I^{({\rm bubble})}_2 }{x_1 x_2 \epsilon }\left(\epsilon  \left((x_2-3) \epsilon ^2-(x_2+10) \epsilon +15\right)-\frac{2 (x_2-1)^2 (\epsilon -1)^2}{x_1^2} \right. 
\\ \nn &+& \left. \frac{2 (x_2-1) \left(\epsilon  \left(x_2 \left(\epsilon ^2+\epsilon +2\right)+3-\epsilon\right)-2\right)}{x_1}-4\right) + I^{({\rm bubble})}_3 \left(\frac{2 (2-x_2) (1-\epsilon )^2}{x_1^3 \epsilon } \right.
\\ \nn &-& \left. \frac{2 (1-\epsilon )}{x_1^2 x_2^2}+\frac{4 \left(\epsilon  x_1^2+(x_1+1)^2\right)}{x_1^3 x_2}+\frac{4 \left(\epsilon ^2+\epsilon +1\right)}{x_1 x_2^2 \epsilon } + \frac{2 x_2 \left(\epsilon ^2+\epsilon +2\right)}{x_1^2}+\frac{2 \left(\epsilon ^2+3 \epsilon +2\right)}{x_1} \right.
\eeqn
\beqn
\nn &+& \left. \frac{4}{x_1 x_2^3}-\frac{4 (2 x_1+1)}{x_1^2 x_2^2}-\frac{4 (x_1+1) \left(\epsilon  x_1^2+x_1+1\right)}{x_1^3 (1-x_3)} - \frac{4 \left(\epsilon ^2-1\right)}{x_1^2 x_2 \epsilon }-\frac{2 \left(\epsilon ^2+1\right)}{x_1 x_2^3 \epsilon } -\frac{2 (1-\epsilon )^2}{x_1^3 x_2 \epsilon } \right. 
\\ &+& \left. \frac{4 \left(x_2^2-x_2+1\right)}{x_2^2 (1-x_3)^2}-\frac{2 \epsilon  (\epsilon +1)}{x_2^2 x_3}-\frac{2 \left(\epsilon ^3+\epsilon +2\right)}{x_1^2 \epsilon }-\frac{2 \left(2 \epsilon ^3+9 \epsilon ^2-2 \epsilon +3\right)}{x_1 x_2 \epsilon }  \right)\, ,
\\ \nn A^{(1)}_5 &=& \frac{2 I^{({\rm box})}_{13} x_3 (x_1 \epsilon +x_3)}{x_2^2}+I^{({\rm bubble})}_0 \left(\frac{4 (2 \epsilon -1)}{x_1^2 x_2 \epsilon }-\frac{2}{x_1^2 x_2}+\frac{2}{x_1^2 (1-x_3)}+\frac{4 \left(2 \epsilon ^2-3 \epsilon +1\right)}{x_1^2 \epsilon } \right.
\\ \nn &-& \left. \frac{2 \left(\epsilon ^2-2 \epsilon \right)}{(1-x_1) x_2}+\frac{4}{x_1 (1-x_3)^2}+\frac{2 \left(\epsilon ^2+\epsilon -2\right)}{x_1(1-x_1)}+\frac{2 (\epsilon -1)}{(1-x_1)^2}\right)
\\ \nn &+& I^{({\rm bubble})}_1 \left(\frac{4 (2 \epsilon -1)}{x_1 x_2^2 \epsilon }+\frac{2 \left(\epsilon ^2-6 \epsilon +2\right)}{x_1 x_2 \epsilon }+\frac{2 \left(\epsilon ^2-2 \epsilon \right)}{(1-x_1) x_2}+\frac{4 (1-\epsilon )}{(1-x_1)x_1} \right.
\\ \nn &-& \left. \frac{2 (\epsilon -1)}{(1-x_1)^2}+\frac{4 \left(2 \epsilon ^2-3 \epsilon +1\right)}{x_2^2 \epsilon }\right) + I^{({\rm bubble})}_3 \left(\frac{4 (1-\epsilon)}{x_1^2 x_2 \epsilon }-\frac{2 (1-2 x_1 \epsilon )}{x_1^2 (1-x_3)} \right. 
\\ &-& \left. \frac{4 \left(2 \epsilon ^2-3 \epsilon +1\right)}{x_1^2 \epsilon }+\frac{2}{x_1 x_2^2}-\frac{4}{x_1 (1-x_3)^2}\right) \, - \left(1 \leftrightarrow 2\right) \, ,
\eeqn
As we can appreciate, the $\epsilon$ dependence of the rational terms is non-trivial which prevents a straightforward extension of the results presented in the main text towards higher orders in $\epsilon$.

\subsection*{$\gamma \to q \bar{q} g$}
In Section \ref{sec:Aqqbarg} we performed a decomposition of $P^{(1)\, , \mu \nu}_{\gamma \to q_1 \bar{q}_2 g_3}$ according to its colour structure. Thus, we can write the coefficients defined in \Eq{EquacionDescomposicionEXACTA1} as
\beqn
A^{(1)}_j \left(\gamma \to q_1 \bar{q}_2 g_3\right) &=& D_A \, A^{(1,D_A)}_j + C_A \, A^{(1,C_A)}_j \, ,
\\ A^{(1,D_A)}_j &=& A^{(1)}_j \left(\gamma \to q_1 \bar{q}_2 \gamma_3\right) \, ,
\eeqn
using the normalization factor given in \Eq{NormalizacionAtoqqbarg}. The terms proportional to $C_A$ are given by
\beqn
\nn A^{(1,C_A)}_1 &=& I^{({\rm bubble})}_0 \left(\frac{2 \epsilon ^2-\epsilon +2}{2 x_1 x_2 \epsilon }-\frac{x_2 (\epsilon -1) \left(2 \epsilon ^2-\epsilon +2\right)}{2 x_1 \epsilon }-\frac{(2 \epsilon +3) (x_2 \epsilon -x_2+\epsilon +1)}{2 (1-x_1)} \right.
\\ \nn &-& \left. \frac{2 \epsilon ^2-\epsilon +2}{x_1 \epsilon }-\frac{1}{x_3}+3 \epsilon +1\right)+I^{({\rm bubble})}_1 \left(\frac{(2 \epsilon +3) (x_2 (\epsilon -1)+\epsilon +1)}{2 (1-x_1)}+\frac{\epsilon  (x_2 \epsilon -1)}{2 x_1} \right.
\\ \nn &+& \left. \frac{2}{x_3}-\frac{\epsilon ^2+6 \epsilon +8}{2}\right) + I^{({\rm box})}_{12} \left(\frac{(1-x_1) (1-x_2)+x_3^3}{4 x_3}-\frac{x_1 x_2}{4 x_3 (1-2 \epsilon )} \right.
\\ &-& \left. \frac{\epsilon}{4}  \left(x_1 x_2+(1-x_3)^2\right)+\frac{5 x_1 x_2}{8 (1-2 \epsilon )}-\frac{1}{8} 9 x_1 x_2\right) \, + \left(1 \leftrightarrow 2\right) \, ,
\\ \nn A^{(1,C_A)}_2 &=& I^{({\rm bubble})}_0 \left(\frac{1-\epsilon ^2}{(1-x_1) x_1}-\frac{\epsilon ^2+3 \epsilon +3}{(1-x_1) x_2}-\frac{2 \epsilon ^2-\epsilon +2}{x_1 x_2}-\frac{2 (1-2 \epsilon )}{(1-x_1) x_3}+\frac{\epsilon -1}{(1-x_1)^2} \right.
\\ \nn &+& \left. \frac{2 (1-\epsilon )}{x_3^2}-\frac{3 \epsilon +1}{x_3}\right)+\frac{I^{({\rm bubble})}_1 }{x_3} \left(\frac{x_2 \left(1-\epsilon ^2\right)}{x_1}+\frac{x_2-(2 x_2+1) \epsilon ^2+(x_2-8) \epsilon}{1-x_1} \right.
\\ \nn &-& \left. \frac{1+\epsilon-\epsilon ^2}{x_1 x_2} - \frac{x_2 (1-\epsilon )}{(1-x_1)^2}+\frac{\epsilon +2}{x_1}+\frac{4 (\epsilon +1)}{x_2}-\frac{4 (1-\epsilon )}{x_3}+\epsilon ^2+5 \epsilon +2\right) 
\eeqn
\beqn
\nn &+& \frac{I^{({\rm box})}_{12} \epsilon  }{1-2 \epsilon } \left(\frac{(1-x_2) x_2 (1-\epsilon )}{x_3^2}-\frac{1-x_2^2 (3\epsilon +1)+x_2 (\epsilon +3)}{2 x_3}+\frac{1}{2} (x_2 (3 \epsilon +1)+5 \epsilon )\right)  
\\ &+& \vphantom{\frac{(1-x_2) x_2 (1-\epsilon )}{x_3^2}} \left(1 \leftrightarrow 2\right) \, ,
\\ \nn A^{(1,C_A)}_3 &=& I^{({\rm bubble})}_0 \left(\frac{\left(x_2^2+1\right) (1-\epsilon )}{2 (1-x_1)^2 x_2}-\frac{x_2 \epsilon +x_2-2 \epsilon }{(1-x_1) x_2^2}+\frac{1-2 \epsilon ^2-3 \epsilon}{2 (1-x_1) x_2}+\frac{2-2 \epsilon ^3+3 \epsilon ^2-3 \epsilon}{x_1 x_2 \epsilon } \right.
\\ \nn &+& \left. \frac{x_2 (1-\epsilon )}{2 (1-x_1) x_1}+\frac{1+\epsilon}{(1-x_1) x_2}-\frac{\epsilon ^2-\epsilon +1}{1-x_1}-\frac{\epsilon ^2+2 \epsilon +2}{x_1}-\frac{1}{(1-x_1)^2} \right.
\\ \nn &+& \left. \frac{x_2(\epsilon +1)-x_2^2(3\epsilon-1) -2 \epsilon }{x_2^2 x_3}-\frac{2 (1-x_2) (1-\epsilon )}{x_2 x_3^2}\right)+I^{({\rm bubble})}_1 \left(\frac{\epsilon ^2-2 \epsilon -3}{2 (1-x_1) x_1 x_2} \right.
\\ \nn &+& \left. \frac{\epsilon +3}{2 (1-x_1)^2 x_2} +\frac{x_2 (\epsilon -1) (x_1 \epsilon -\epsilon +1)}{2 (1-x_1)^2 x_1}+\frac{2 x_1 (1-\epsilon )}{(1-x_1) x_3^2}+\frac{2-(1-x_1) x_1 (3 \epsilon +1)}{(1-x_1)^2 x_3} \right.
\\ \nn &+& \left. \frac{\epsilon ^2+\epsilon +2}{x_1}+\frac{(1-\epsilon )^2}{1-x_1}+\frac{1}{(1-x_1)^2}\right)+I^{({\rm bubble})}_2 \left(\frac{\epsilon ^2+1}{2 x_1 x_2}+\frac{\epsilon ^2+5 \epsilon +2}{2 x_1} \right.
\\ \nn &+& \left. \frac{2 (1-x_2) (1-\epsilon)}{x_2 x_3^2}-\frac{1-(3 x_2+1) \epsilon -x_2}{x_2 x_3}\right) + I^{({\rm box})}_{12} \left(\frac{(1-x_2)^2 (\epsilon-1 ) \epsilon }{x_3^2 (1-2 \epsilon )} \right.
\\ &+& \left. \frac{(1-x_2) \epsilon  (3-(3 x_2+1) \epsilon -x_2)}{2 x_3 (1-2 \epsilon )}-\epsilon +1\right) \, ,
\\ \nn A^{(1,C_A)}_5 &=& \frac{I^{({\rm bubble})}_0 }{x_3}\left(\frac{x_2 \left(2-\epsilon ^2-\epsilon\right)}{(1-x_1) x_1}+\frac{x_2 (1-\epsilon )}{(1-x_1)^2}-\frac{\epsilon ^2-3 \epsilon +3}{1-x_1}-\frac{\epsilon +2}{x_1}\right)-\frac{I^{({\rm bubble})}_1 }{x_3} \left(\frac{2 \epsilon }{x_2} \right.
\\ \nn &+& \left. \frac{x_3 (\epsilon +2)}{x_1 x_2}+\frac{x_2 (1-\epsilon )}{(1-x_1)^2}+\frac{2 x_2 (1-\epsilon )}{x_1}-\frac{(3-2 x_2) (1-\epsilon )}{1-x_1}-\frac{2 \epsilon }{x_1}+\frac{x_3 \epsilon ^2}{(1-x_1) x_2}\right) 
\\ &+& \frac{I^{({\rm box})}_{12} x_2 \epsilon }{x_3} \, - \left(1 \leftrightarrow 2\right)\, .
\eeqn

%%%%%%%%%%%%%%%%%%%%%%%%%%%%%%%%%%%%%%%%%%%%%%%%%%%%%%%%%%%%%%%%%%%%%%%%%%%%%%%%%%%%%%%%

\end{document}